\documentclass[iop]{emulateapj}
\usepackage{graphicx}
\usepackage{subfigure}
\usepackage{multirow}
\usepackage{amssymb}
\usepackage{natbib}
\usepackage{rotating}
\usepackage[bookmarks=false, colorlinks=true, citecolor=blue, backref=true]{hyperref}
\usepackage{apjfonts}
\shorttitle{Turbulent molecular gas and star formation in Stephan's Quintet}
\shortauthors{Guillard et al.}

\begin{document}


\title{Turbulent molecular gas and star formation in the shocked intergalactic medium of Stephan's Quintet}



 \author{P. Guillard\altaffilmark{1,2}}, \author{F. Boulanger\altaffilmark{2}}, \author{G. Pineau des For\^ets\altaffilmark{2,6}}, 
\author{E. Falgarone\altaffilmark{6}}, \author{A. Gusdorf\altaffilmark{3,4}}, 
 \author{M. E. Cluver\altaffilmark{1}}, \author{P. N. Appleton\altaffilmark{5}}, \author{U. Lisenfeld\altaffilmark{7,8,1}}, \author{P.-A. Duc\altaffilmark{9}}, \author{P. M. Ogle\altaffilmark{1}}, 
  \author{C. K. Xu\altaffilmark{10}}

\altaffiltext{1}{\textit{Spitzer} Science Center (SSC), California Institute of Technology, MC 220-6, Pasadena, CA-91125}
\altaffiltext{2}{Institut d'Astrophysique Spatiale (IAS), UMR 8617, CNRS, Universit\'e Paris-Sud 11, B\^atiment 121, 91405 Orsay Cedex, France}
\altaffiltext{3}{Max Planck Institut f\"{u}r Radioastronomie, Auf dem H\"ugel 69, 53121 Bonn, Germany}
\altaffiltext{4}{Laboratoire Univers et Th\'eories (LUTH), UMR 8102 CNRS, Observatoire de Paris, Universit\'e Paris Diderot, 5 Place Jules Janssen, 92190 Meudon, France}
\altaffiltext{5}{NASA \emph{Herschel} Science Center (NHSC), California Institute of Technology, Mail code 100-22, Pasadena, CA 91125, USA}
\altaffiltext{6}{ENS, LERMA, UMR 8112, CNRS, Observatoire de Paris, 61 Avenue de l'Observatoire, 75014 Paris, France}
\altaffiltext{7}{Departamento de F\'isica Te\'orica y del Cosmos, Universidad de Granada, Spain}
\altaffiltext{8}{Instituto de Astrof\'isica de Andaluc\'ia, CSIC, Apdo. 3004, 18080 Granada, Spain}
\altaffiltext{9}{AIM, Unit\'e Mixte de Recherche CEA-CNRS, Universit\'e Paris VII, UMR 7158, France}
\altaffiltext{10}{Infrared Processing and Analysis Center (IPAC), JPL, Pasadena, USA}

\begin{abstract}
The Stephan's Quintet (hereafter SQ) is a template source to study the
impact of galaxies interaction on the physical state and energetics of their gas. 
We report on IRAM single-dish CO observations of the SQ compact group of galaxies. These
observations follow up the \textit{Spitzer} discovery of bright mid-IR H$_2$  rotational
line emission ($L(\rm H_2) \approx 10^{35}$~W) from warm ($10^{2-3} ~$K) molecular
gas, associated with  a 30 kpc long shock between a galaxy, NGC 7318b, and NGC
7319's tidal arm. We detect CO(1-0), (2-1) and (3-2) line emission in the inter-galactic medium (IGM) with complex profiles,
spanning a velocity range of  $\approx 1000$~km~s$^{-1}$. The spectra exhibit the
pre-shock recession velocities of the two colliding gas systems ($5700$ and $6700$~km~s$^{-1}$), but also
intermediate velocities. This shows that much of the molecular gas has formed out of diffuse gas accelerated by the
 galaxy-tidal arm collision. CO emission is also detected in a bridge feature that connects the shock
to the Seyfert member of the group, NGC 7319, and in the northern star forming region, SQ-A, where
a new velocity component is identified at $\rm 6900 ~km s^{-1}$, in addition to the two velocity components
already known. Assuming a Galactic CO(1-0) emission to H$_2$  mass conversion factor, a total
H$_2$  mass of  $\approx 5\times10^9\,$M$_{\odot}$ is detected in the shock.  
The ratio between the warm H$_2$  mass derived from \textit{Spitzer} spectroscopy, and the H$_2$  mass 
derived from CO fluxes is $\approx 0.3$ in the IGM of SQ, which is 10--100 times higher than
in star-forming galaxies. The molecular gas carries a large fraction of the gas kinetic energy
involved in the collision, meaning that this energy has not been thermalized yet.
The kinetic energy of the H$_2$ gas derived from CO observations is comparable to that of the warm H$_2$
gas from \textit{Spitzer} spectroscopy, and a factor 
$\approx 5$ greater than the thermal energy of the hot plasma heated by the collision.  
In the shock and bridge regions, the ratio of the PAH-to-CO surface luminosities, commonly used to measure
the star formation efficiency of the H$_2$ gas, is lower (up to a factor 75)  than the observed values in star-forming galaxies.
We suggest that turbulence fed by the galaxy-tidal arm collision
maintains a high heating rate within the H$_2$  gas. This interpretation
implies that the velocity dispersion on the scale of giant molecular clouds 
in SQ is one order of magnitude larger than the Galactic value. The high amplitude of turbulence may explain why this gas is not forming stars efficiently.

\end{abstract}


%

\section{Introduction}

Galaxy interactions represent an important agent of galaxy evolution, involving the injection of huge amounts of kinetic energy in the interstellar medium. Galaxy collisions  or interactions of galaxies with the inter-galactic medium (henceforth IGM) may trigger bursts of star formation detected in the  infra-red (IR)  \citep[e.g. ][]{Joseph1985, Harwit1987, Kim1995a, Xu1999}. 
Star formation proceeds where and when the kinetic energy of the gas is being radiated, allowing gas to cool and condense. To make 
headway in our understanding of galaxy evolution and, in particular,  the star formation history of the universe, it is crucial to understand how this kinetic energy is dissipated, and how this dissipation affects the passage from molecular gas to stars.

Stephan's Quintet (Hickson Compact Group \object{HCG 92}), hereafter SQ, is an ideal laboratory to study these physical processes. SQ is a group of five interacting galaxies, with a complex dynamical history \citep[see][for a review and references therein]{Xu2006}. The most striking feature of the group is that a galaxy-scale ($\approx 15 \times 35$~kpc$^{2}$) shock is created by an intruding galaxy, NGC~7318b which is colliding into the intra-group gas at a relative velocity of $\approx \! 1\,000$~km~s$^{-1}$ (see the background image in Figure~\ref{fig:CO_beams_on_HST}). This IGM material  is likely to be gas that has been tidally stripped from NGC~7319 by a past interaction with another galaxy, NGC 7320c \citep[see][for a plausible dynamical scenario]{Moles1997, Sulentic2001}. These ideas have been further explored recently with numerical simulations.  Both N-body \citep{Renaud2010} and hydrodynamical models \citep{Hwang2011} of the Quintet suggest a sequence of  serial collisions between various group members culminating in the current high-speed shock-heating of gas
as NGC~7318b strikes the pre-existing tidal debris from NGC~7319.

A ridge of X-ray  \citep{Trinchieri2003, Trinchieri2005} and strong radio synchrotron \citep[e.g., ][]{Sulentic2001, Williams2002} emission from the hot ($5-7\times 10^6\,$K) post-shock plasma is associated with the group-wide shock. Excitation diagnostics from optical \citep{Xu2003} and mid-IR \citep{Cluver2010} emission lines  also confirm the presence of shocked gas in the IGM. \textit{Spitzer} mid-IR observations reveal that the mid-IR spectrum in the shock structure is dominated by the rotational line emission of molecular hydrogen, $\rm H_2$ \citep{Appleton2006, Cluver2010}. This denotes the presence of  large amounts ($\approx 10^9\,$M$_{\odot}$) of  warm ($10^2-10^3\,$K) $\rm H_2$, co-spatial with the X-ray emitting plasma of the group-wide shock. The $\rm H_2$ emission is extended (see the $\rm H_2$ red contours on Fig~\ref{fig:CO_beams_on_HST}) along the ridge feature, towards the Seyfert galaxy NGC~7319 (in the so-called \textit{bridge}), as well as in the northern intergalactic starburst SQ-A \citep{Xu1999}.   Over the main shock area observed by \textit{Spitzer}, the $\rm H_2$ luminosity ($L({\rm H}_2) \approx 2.6 \times 10^8 \,$L$_{\odot}$, summed over the 0-0S(0) to S(5) lines) is larger by a factor of three than that of the X-ray emission (integrated over $0.001-10\,$keV) from the same region. 
With the high resolution ($\mathcal{R}=600$) of  \textit{Spitzer} IRS, the $17\,\mu$m H$_2$ S(1)  line is resolved with a FWHM of 870 km~s$^{-1}$ \citep{Appleton2006}.

\citet{Guillard2009} proposed a model for the formation of molecular gas in the ridge, considering the collision of multiphase gas. In their scenario, $\rm H_2$ is formed from the cooling of shocked atomic gas. The shock velocity depends on the pre-shock gas density: it is high ($\approx 600-800\,$km~s$^{-1}$) in the tenuous ($n_{\rm H} < 0.02\,$cm$^{-3}$) gas, producing the X-ray emission in the ridge, and much lower in the dense gas. Gas at pre-shock densities  $n_{\rm H} > 0.2\,$cm$^{-3}$ experience shock velocities small enough ($V_{\rm s} < 200$~km~s$^{-1}$) to retain most of its dust, cool, and  become molecular within a few million years. 
We proposed that the $\rm H_2$ emission is associated with the dissipation of the kinetic energy of the galaxy collision at sub-parsec scales, through supersonic turbulence in the  molecular gas. The mid-IR spectral energy distribution (SED) of the $\rm H_2$ emission can be explained if the dense ($n_{\rm H} > 10^3\,$cm$^{-3}$) molecular gas is continuously processed by low-velocity ($5-20$~km~s$^{-1}$) MHD shocks.

Since the mid-IR rotational $\rm H_2$ line emission only samples warm molecular gas with temperatures $\approx 10^2-10^3\,$K, our census of the molecular gas in the SQ shock is still incomplete. The discovery of powerful mid-IR $\rm H_2$ line emission in SQ led us to search for CO line emission associated with the warm $\rm H_2$ in the shock. In galaxies, the warm H$_2$ is a small fraction ($0.4-4$~\%) of the total mass of molecular gas \citep{Roussel2007}, which is too cold ($< 100$~K) to be seen in mid-IR H$_2$ emission, and usually traced with CO. 
This paper reports  on the detection of CO line emission in the IGM of the SQ group, and investigates the physical and dynamical state of the molecular gas in SQ, as well as the connection between shock and star formation in the group.

After reviewing the past CO observations of SQ (Sect.~\ref{sec:previousCOobs}), we present technical details about our observations with the IRAM 30-meter  and the APEX telescopes (Sect.~\ref{sec:CO_obs}). Then we describe the kinematics  (Sect.~\ref{sec:kinetics_CO}), the distribution and mass (Sect.~\ref{sec:distribution_mass_CO}), the excitation and energetics (Sect.~\ref{sec:excit_nrj}) of  the  molecular gas in the IGM of the group. Sect.~\ref{sec:discussion} discusses the origin of the CO gas, its kinematics, and the efficiency of star formation in the group. Conclusions and final remarks are given in Sect.~\ref{sec:conclusion}.

Throughout this paper we assume the distance to the SQ group to be 94~Mpc (with a Hubble
constant of 70 km~s$^{-1}$~Mpc$^{-2}$) and a systemic velocity for the
group as a whole of $6\,600$~km~s$^{-1}$. At this distance, $10\ \rm arcsec=4.56\,$kpc.

\section{Previous CO observations in the SQ group}
\label{sec:previousCOobs}

Before the discovery of warm $\rm H_2$ in the IGM of the SQ group, previous observations were mostly concentrated on the Seyfert galaxy NGC~7319, on the southern tidal features, or on the northern starburst region (SQ-A). 
The first millimetre CO observations in SQ detected cold molecular gas in NGC 7319 \citep{Yun1997, Verdes-Montenegro1998, Leon1998}, which contains $4.8 \times 10^9$ M$_{\odot}$ \citet{Smith2001a}. 
Using the BIMA\footnote{Berkeley Illinois Maryland Association, \url{http://bima.astro.umd.edu/bima.html}} interferometer, \citet{Gao2000, Petitpas2005} showed that $3/4$ of the CO gas in NGC~7319 is lying outside its nucleus, mostly in an extended northern region (see green contours on Figure~\ref{fig:CO_beams_on_HST}).

Molecular gas outside galactic disks, associated with the IGM starburst SQ-A, was first detected by \citet{Gao2000} using BIMA. Then, single dish observations by \citet{Smith2001a} and \citet{Lisenfeld2002} confirmed this result. 
Two velocity components were detected, one centred at
$\sim 6000$~km~s$^{-1}$, and the other at $\sim 6700$~km~s$^{-1}$. These velocities match the redshifts of
the two H$\,${\sc i} gas systems found in the same region \citep{Sulentic2001, Williams2002}, which correspond to the intruder galaxy, NGC~7318b, and NGC~7319's tidal tail in the IGM, respectively.
\citet{Lisenfeld2002} found that there is more molecular gas ($3.1 \times 10^9$~M$_{\odot}$)
in SQ-A than  H$\,${\sc i} gas ($1.6 \times 10^9$~M$_{\odot}$). 

Past observations with the BIMA interferometer did not lead to a CO detection in the shock itself \citep{Gao2000}. The  single-dish observations by \citet{Lisenfeld2002} using the IRAM 30m telescope partially overlap the shock structure  and show a $3\sigma$ detection of CO in the northern region of the X-ray ridge, but their observations do not probe the entire shock region. We note that they include this emission in their SQ-A CO flux.  \citet{Gao2000, Smith2001a, Petitpas2005} reported CO emission close to NGC~7318b nucleus and from several other regions in the group. 

Another interesting feature of SQ is the presence of a CO-rich \citep[$7 \times 10^8$~M$_{\odot}$]{Braine2001, Lisenfeld2002, Lisenfeld2004} region, SQ-B, located within the young H{\sc i} tidal tail \citep[see][]{Sulentic2001}, to the south of NGC~7319 (outside the field of view of Figure~\ref{fig:CO_beams_on_HST}). This region is also bright in the mid-IR \citep{Xu1999, Guillard2010}, H$\alpha$ \citep{Arp1973a} and UV \citep{Xu2005}. The $\rm H_2$-to-H$\,${\sc i} gas mass ratio of 0.5 indicates that SQ-B is a tidal dwarf galaxy (TDG) candidate, i.e. a small galaxy which is in the process of formation  from the material of the NGC~7319's tidal tail \citep{Boquien2009}. 
The gas metallicity in SQ-A and SQ-B is slightly higher than solar \citep{Xu2003, Lisenfeld2004}. This is a strong indication that the gas in these regions has been pulled out from the inner part of a galaxy disk (or maybe several disks) by tidal interactions. TDGs will not be further discussed, as we mostly concentrate on the shock, the bridge and SQ-A in this paper.

\section{CO observations}
\label{sec:CO_obs}

\subsection{CO(1-0) and (2-1) with the IRAM 30m telescope}
\label{subsec:CO_obs_emir}

\begin{figure*}[h!]
   \centering
    \includegraphics[width=\textwidth]{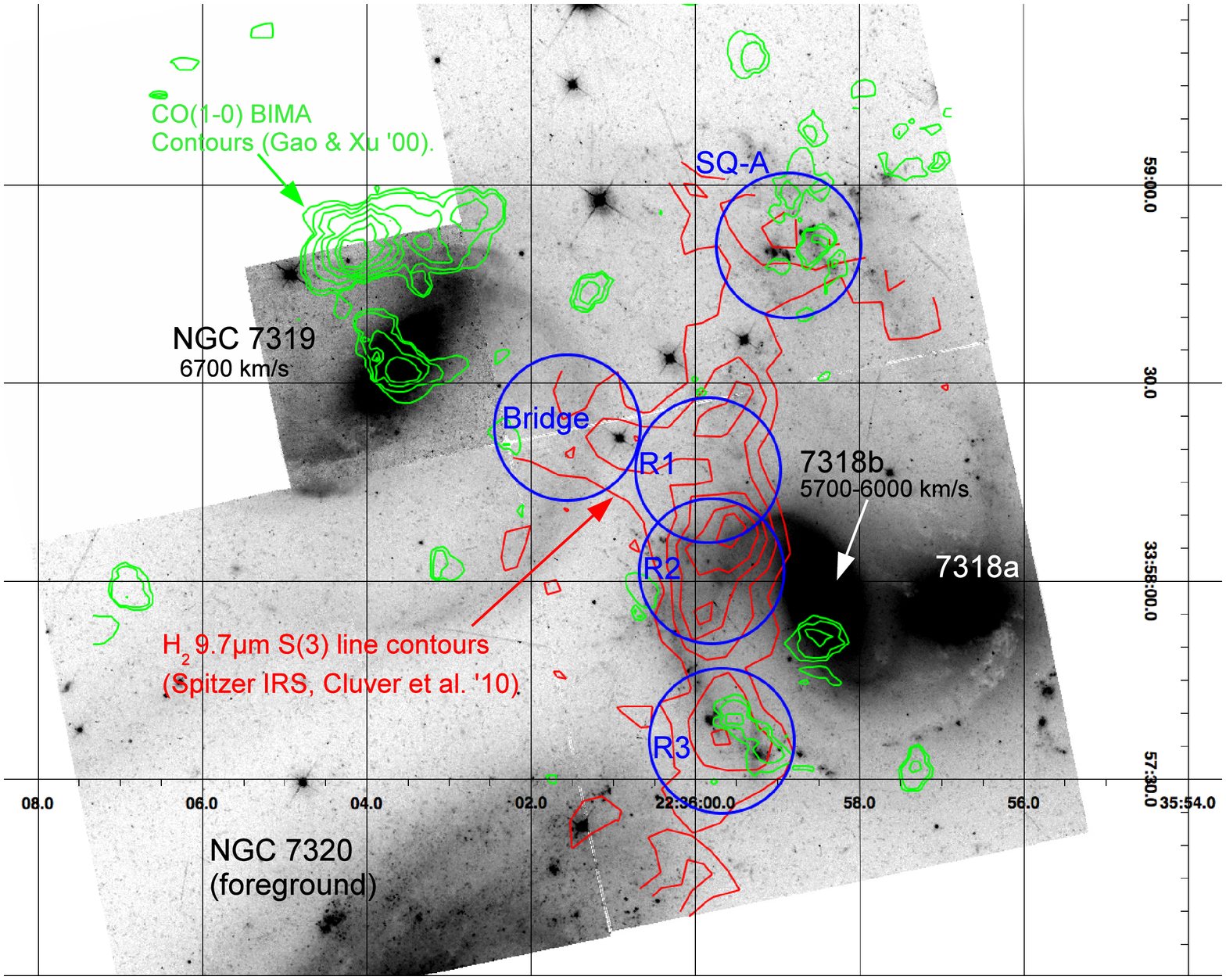}
      \caption{CO and $\rm H_2$ observations in the Stephan's Quintet group. The grey-scale background image is  a \textit{HST} WFC2  $V$-band image. The blue circles (22 arcsec in diameter) show the IRAM 30m CO(1-0) half-power beams where the observations presented in this paper have been taken.  The corresponding spectra are shown on Figure~\ref{fig:SQ_CO_spectra}. CO emission is detected at all these positions. 
      The intruder galaxy, NGC~7318b, is colliding with intra-group gas (tidal debris from NGC~7319) at a relative velocity of $\sim 900\,$km~s$^{-1}$.  The CO(1-0) green contours from BIMA observations by \citet{Gao2000} correspond to levels of 2.4, 3.0, 4.0, 5.0, 6.0, 7.0, and 8$\sigma$ ($1 \sigma = 0.7$ Jy km/s). No CO was detected in the ridge because of limited sensitivity and bandwidth.  Also note that the two $\sim 3-4 \sigma$ CO features in the south were just outside (but near the edge) of the BIMA primary beam. The red contours are the $\rm H \ _2$ $9.7\,\mu$m S(3) line  contours from \textit{Spitzer IRS} spectral mapping by \citet{Cluver2010}, which clearly shows the kpc-scale shock produced by the galaxy collision. The $\rm H \ _2$ contours are [1.0, 1.6, 2.3, 2.9, 3.6, 4.2, 4.9]$\times 10^{-9}$~W~m$^{-2}$~sr$^{-1}$. }
       \label{fig:CO_beams_on_HST}
   \end{figure*}

\begin{deluxetable}{lccccc} 
\tablecolumns{6}
\tablecaption{CO observations log}
\tablehead{
\colhead{ \multirow{2}*{Position}} &  \multirow{2}*{RA (J2000)} & \multirow{2}*{DEC (J2000)} & \multicolumn{3}{c}{ integration time [min]} \\
 & & & \colhead{CO(1-0)} & \colhead{CO(2-1)} & \colhead{CO(3-2)} 
}
\startdata
SQ-A\tablenotemark{a}	 & 22:35:58.88  & +33:58:50.69   & 122 & 58   & 66 \\
Ridge 1   						     & 22:35:59.85  & +33:58:16.55   & 311 & 186 & 80 \\
Ridge 2  						         & 22:35:59.85  &  +33:58:04.00  & 318 & 249 & \nodata \\
Ridge 3   							 & 22:35:59.73  & +33:57:37.16   & 86    & 41   & \nodata \\
Bridge\tablenotemark{b} & 22:36:01.56  &  +33:58:23.30  & 175 & 57    & \nodata 
\enddata
\tablenotetext{a}{Starburst region at the northern tip of the SQ ridge}
\tablenotetext{b}{Bridge feature that connects the ridge to the AGN NGC~7319}
\tablecomments{Pointing positions and useful integration times of the CO spectra presented in this paper. The CO(1-0) and (2-1) lines were observed with the IRAM 30m telescope, and the CO(3-2) with the APEX telescope.}
\label{table:CO_obs_log}
\end{deluxetable}

The observations were made on June 28-30, 2009 at the IRAM 30~m telescope  at Pico Veleta, Spain. We used the  heterodyne receiver, \textit{EMIR} \footnote{For details about the Eight MIxer Receiver instrument, see \url{http://www.iram.es/IRAMES/mainWiki/EmirforAstronomers}}, commissioned in March-April 2009.  \textit{EMIR} provides a bandwidth of 4~GHz in each of the two orthogonal linear polarizations for the 3, 2, 1.3 and 0.9~mm atmospheric windows. The four \textit{EMIR} bands are designated as E090, E150, E230, and E330 according to their approximate center frequencies in GHz. Given the wide range of CO velocities ($6000-6700\,$km~s$^{-1}$) exhibited by previous observations, and the broad linewidth of the S(1) $\rm H_2$ line in the shock,  this increase in bandwith was crucial to our study. 

We observed the CO(1-0) and CO(2-1) lines at 115.27~GHz and 230.54~GHz by connecting the third and fourth parts of the WILMA backend to the horizontal and vertical polarizations of the E230 band.  We focus this paper on the CO(1-0) results, since we did not sample the (1-0) beam with the (2-1) observations. To increase the redundancy of the CO(1-0) data, we used in parallel the 4~MHz filterbank with both polarizations connected to the E090 band. The receivers were tuned to 112.86~GHz and 225.72~GHz, which corresponds to a recession velocity of 6400~km~s$^{-1}$. The observations were done in wobbler switching mode, with a wobbler throw of 120'' in azimuthal direction.
This observation mode provides accurate background subtraction and good quality baselines, which is critical for measuring broad lines.
 Pointing and focus were monitored every 2h in stable conditions and every 1h during sunrise. The pointing accuracy was $\sim 3''$ and the system temperatures were $\approx 150-200$~K at 115 GHz on the $T_{A}^{*}$ scale. The  forward efficiency of the telescope was 0.94 and 0.91 at 115 and 230 GHz and the main beam efficiency was 0.78 and 0.58, respectively.
The half-power beam size is $22''$ at 115 GHz and $11''$ at 230 GHz.

The data was reduced with the IRAM GILDAS-CLASS90\footnote{Grenoble Image and Line Data Analysis System, \url{http://www.iram.fr/IRAMFR/GILDAS/}} software package (version of December 2010). The lowest quality spectra (typically with an RMS noise $\sigma \gtrsim 30$~mK for a 3~min on-source integration)  were rejected. We subtracted  linear  baselines. 
The final spectra presented in this paper are smoothed (with a hanning window) to a velocity resolution of 40~km~s$^{-1}$.
We experienced problems with the vertical polarization. These spectra, which showed spurious features, ripples and curved baselines, have been  rejected for this analysis. 

We observed five different positions in the center of the SQ group. The blue circles on  Figure~\ref{fig:CO_beams_on_HST} show the CO(1-0) half-power beams ($22''$ in diameter)  overlaid on a \textit{HST WFPC2} image of SQ. 
The three positions in the ridge are designated as R1, R2 and R3, from North to South. The two other positions are centred on the SQ-A star forming region and on the middle of the $\rm H_2$-bridge feature in between NGC~7319 and the main shock.
The red lines show the  contours of the $\rm H_2$ 0-0S(3) line emission detected by \textit{Spitzer} \citep{Cluver2010}. 
Note that the half power beam width at 115~GHz matches very well the width of the $\rm H_2$ emission in the ridge.
Table~\ref{table:CO_obs_log} lists the coordinates of these positions and the total ON+OFF time spent at each position. A total of $\approx 20$ hours has been spent for ON+OFF observations towards SQ.

\begin{figure*}
   \centering
    \includegraphics[width=\textwidth]{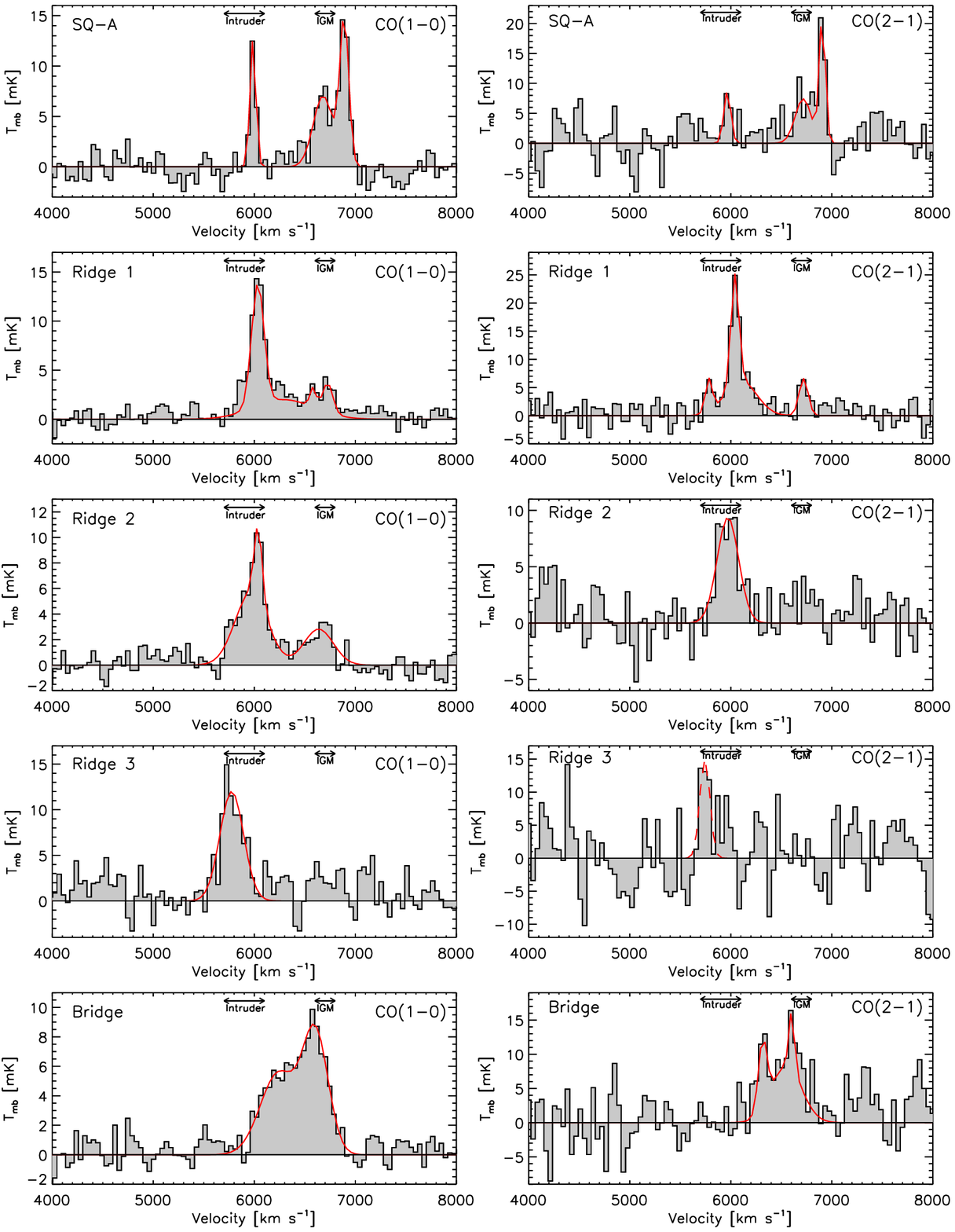}
      \caption{CO(1-0) and CO(2-1) spectra (left and right columns) observed with the IRAM 30m telescope at the five positions observed in the IGM of the SQ group (see Figure~\ref{fig:CO_beams_on_HST}). CO emission with multiple velocity components is detected at all positions: in the shock, in SQ-A, and in the bridge regions. The red line shows the result of the line fitting (see Tables \ref{table_CO10_results}  and \ref{table_CO21_results} for the decomposition in multiple Gaussian components for each transition). The spectra have been smoothed to a spectral resolution of 40~km~s$^{-1}$. Note that the CO(1-0) and CO(2-1) beams do not match, and that the detection of the CO(2-1) line in the Ridge~3 position is tentative (in this case the Gaussian fitting --dashed line-- is not reliable). The arrows indicate the pre-shock H{\sc i} velocities for the intruder galaxy and the intra-group medium.}
       \label{fig:SQ_CO_spectra}
   \end{figure*}

\begin{deluxetable*}{lcccccc}
 \tablecaption{CO(1-0) observations with the IRAM 30m telescope: results from Gaussian decompositions\label{table_CO10_results}}
 \tablecolumns{7}
 \tablewidth{0.8\textwidth}
 \tablehead{
 \multirow{2}*{Target region}  & \colhead{$\rm v _{\rm CO(1-0)}$ }  & \colhead{$\Delta \rm v_{\rm CO(1-0)}$  }  &	\colhead{$\mathcal{I}_{\rm CO(1-0)}$ } & \colhead{$M(\rm H_2)$  \tablenotemark{a}} & \colhead{$E_{\rm kin}$ \tablenotemark{b}} & \colhead{ \multirow{2}*{$M(\rm H_2^{\rm w}) / M(\rm H_2^{\rm c})$  \tablenotemark{c}}}  \\	
										& \colhead{[km s$^{-1}$]} &   \colhead{[km s$^{-1}$]} &	  \colhead{[K km s$^{-1}$]} &  \colhead{[$\times 10^8$~M$_{\odot}$]} & \colhead{[ergs]} & 
}
\startdata
 \multirow{3}*{SQ-A}	    	  & $5987 \pm 14$   &  $69 \pm 8$       & $0.9 \pm 0.1$ & $ 3.1 \pm 0.5$   &  $7.8 \pm 3.1 \ E54$   &   \\
	  										     & $6681 \pm 22$    & $230 \pm 35$    & $1.7 \pm 0.3$ & $ 5.7 \pm 1.1 $  &  $1.6 \pm 0.8 \ E56$   &   \\
	  										     & $6891 \pm 11$    & $104 \pm 12$    & $1.6 \pm 0.3$ & $ 5.3 \pm 0.9 $  &  $3.1 \pm 1.2 \ E55$    &  \\
\tableline
SQ-A (sum) \tablenotemark{d}	   & &  	                                         &	$4.2 \pm 0.4$ & $14.1 \pm 1.5$  &  $2.0 \pm 0.8 \ E56$    &   $ 0.9 \pm 0.2  $  \\
\tableline
 \multirow{3}*{Ridge~1}     & $6034 \pm 11$   & $144 \pm 10$    & $1.9 \pm 0.2$     & $6.3 \pm 0.8$     & $7.1 \pm 1.8\ E55$     &    \\
 											  & $6288 \pm 22$   & $700 \pm 60$    & $1.5 \pm 0.2$     & $5.0 \pm 0.7$     & $1.3 \pm 0.4\ E57$     &      \\
 											  & $6581 \pm 20$   & $71 \pm 8$        & $0.15 \pm 0.02$ & $0.50 \pm 0.08$ & $1.3 \pm 0.5\ E54$     &      \\
  											  & $6727 \pm 15$   & $118\pm 13$     & $0.4 \pm 0.06$   & $1.2 \pm 0.2$     & $9.4 \pm 0.4\ E54$     &    \\ 											 
\tableline
Ridge~1 (tot) \tablenotemark{d} 							  & &  			     &	$4.0 \pm 0.3$ & $13.0 \pm 1.4$ &  $1.4 \pm 0.4\ E57$    &   $ 0.26 \pm  0.03 $  \\
\tableline
\multirow{3}*{Ridge 2} 	& $5970 \pm 28$ & $377 \pm 36$      & $2.3 \pm 0.3$      & $7.5 \pm 1.1$ & $5.8 \pm 1.9\ E56$      &    \\
											& $6038 \pm 12$ & $93 \pm 10$        & $0.56 \pm 0.08$  & $1.9 \pm 0.3$ & $8.6 \pm 3.2\ E54$     &       \\  
											& $6637 \pm 35$ & $342 \pm 28$      & $1.0 \pm 0.1$      & $3.4 \pm 0.4$ & $2.1 \pm 0.6\ E56$      &    \\									
\tableline
Ridge 2 (tot) \tablenotemark{d}  	& &  									     & $3.87 \pm 0.32$  & $12.8 \pm 1.2$ & $8.0 \pm 2.0\ E56$     &   $ 0.33 \pm 0.07  $  \\
\tableline	
	Ridge 3                            & $5777 \pm 25$ & $276 \pm 19$    & $3.5 \pm 0.4$      & $11.6 \pm 1.4 $ & $4.8 \pm 1.2\ E56$      &   $ 0.10 \pm 0.02  $ \\							  
\tableline
\multirow{2}*{Bridge}     & $6239 \pm 21$   & $423 \pm 24$   & 	$2.5 \pm 0.3$ & $8.2 \pm 0.9 $    & $7.9 \pm 1.8\ E56$      &   \\
										  & $6605 \pm 20$   & $298 \pm 19$   & $2.6 \pm 0.4$  & $8.5 \pm 1.0$     & $4.1 \pm 1.0\ E56$     &    \\	
\tableline
Bridge (tot) \tablenotemark{d}  & &  									           &	$5.1 \pm 0.5$ & $16.7 \pm 1.4$ & $1.2 \pm 0.2\ E57$      &    $ 0.55 \pm 0.09  $
\enddata
\tablecomments{Observational results of the IRAM 30m EMIR CO(1-0) observations. This table gathers the results from the fitting of the CO(1-0) spectra in multiple Gaussian velocity components (see Figure $\ $~\ref{fig:SQ_CO_spectra}). The first three columns list the central velocity, FWHM, and intensity of the Gaussian components. Results are given for each velocity components, as well as for the sum of the components. The velocity component at 6900 km s$^{-1}$ in SQ-A is a newly discovered feature (see text for details).  The aperture of the beam is $1.13 \times (22\  \rm arcsec)\ ^2 = 547 \ \rm arcsec \,^2$.}
\tablenotetext{a}{$\rm H_2$ gas masses calculated using the Galactic conversion factor of $N({\rm H\  _{2}})/ I_{\rm CO} = 2 \times 10^{20}$~cm$^{-2}$ [K~km~s$^{-1}$]$^{-1}$}
\tablenotetext{b}{Kinetic energy in random motions, calculated as $E\ _{\rm kin} = 3/2 M({\rm H_{2}}) \sigma _{\rm CO} ^2$ (see Sect.~\ref{subsec:kin_energy}).}
\tablenotetext{c}{Ratio between the $\rm H\; _2$ mass derived from the \textit{Spitzer} spectral mapping of the mid-IR $\rm H\; _2$ lines in the group to the H$\; _2$ mass derived from the CO(1-0) line intensity at the different positions in the IGM of SQ.  }
\tablenotetext{d}{Sum over all velocity components}
\end{deluxetable*}

\begin{deluxetable*}{lcccccc}
 \tablecaption{CO(1-0) observations with the IRAM 30m telescope: results from integrations of the spectra over velocity ranges\label{table_CO10_results_int}}
 \tablecolumns{7}
 \tablewidth{0.8\textwidth}
 \tablehead{
 \multirow{2}*{Target region}  & \colhead{Velocity range}  & \colhead{$ \langle \rm v \rangle$  }  &	\colhead{$\mathcal{I}_{\rm CO(1-0)}$ } & \colhead{$M(\rm H_2)$  \tablenotemark{a}} & \colhead{$E_{\rm kin}$ \tablenotemark{b}} & \colhead{ \multirow{2}*{$M(\rm H_2^{\rm w}) / M(\rm H_2^{\rm c})$  \tablenotemark{c}}}  \\	
										& \colhead{$\rm v_{min} - v_{max}$ [km s$^{-1}$]} &   \colhead{[km s$^{-1}$]} &	  \colhead{[K km s$^{-1}$]} &  \colhead{[$\times 10^8$~M$_{\odot}$]} & \colhead{[ergs]} & 
}
\startdata
 \multirow{4}*{SQ-A}	    	  & $5900-6200$   & $5984$ & $0.9 \pm 0.1$   & $ 3.0 \pm 0.2 $  &  $3.2 \pm 0.8 \ E54$   &   \\
	  										  & $6200-6500$   & $6375$ & $0.4 \pm 0.05$ & $ 1.3 \pm 0.9 $  &  $3.5 \pm 0.8 \ E55$   &   \\
	  										  & $6500-6800$   & $6676$ & $1.5 \pm 0.2$   & $ 4.8 \pm 0.5 $  &  $7.4 \pm 1.6 \ E55$    &  \\
	  										  & $6800-7100$   & $6882$ & $1.7 \pm 0.2$   & $ 5.6 \pm 0.6 $  &  $1.7 \pm 0.5 \ E55$    &  \\
\tableline
SQ-A (sum) \tablenotemark{d}	   & &  	                           & $4.5 \pm 0.3$  & $ 14.7 \pm 1.2$  &  $1.3 \pm 0.2 \ E56$    &   $0.8 \pm 0.2  $  \\
\tableline
 \multirow{5}*{Ridge~1}      & $5500-5900$   & $5831$  & $0.4 \pm 0.05$ & $1.3 \pm 0.3$   & $2.2 \pm 0.6\ E55$     &    \\
 											 & $5900-6200$   & $6041$  & $2.5 \pm 0.3$   & $8.3 \pm 1.0$   & $1.1 \pm 0.4\ E56$     &      \\
 											 & $6200-6500$   & $6325$  & $0.8 \pm 0.1$   & $2.5 \pm 0.1$   & $5.5 \pm 0.3\ E55$     &      \\
 											 & $6500-6800$   & $6656$  & $0.9 \pm 0.1$   & $2.9 \pm 0.1$   & $6.4 \pm 0.3\ E55$     &      \\
  											 & $6800-7500$   & $7039$  & $0.5 \pm 0.05$ & $1.6 \pm 0.2$   & $1.0 \pm 0.4\ E56$     &    \\ 											 
\tableline
Ridge~1 (sum) \tablenotemark{d} 							  & &  & $5.2 \pm 0.4$ & $16.6 \pm 1.5$   &  $3.5 \pm 0.8\ E56$    &   $ 0.20  \pm 0.03  $  \\
\tableline
\multirow{3}*{Ridge 2} 	    & $5700-6200$ & $5971$     & $2.6 \pm 0.3$  & $8.6 \pm 1.2$ & $3.5 \pm 0.8\ E56$      &    \\
											& $6200-6500$ & $6346$     & $0.4 \pm 0.05$& $1.5 \pm 0.3$ & $3.8 \pm 0.9\ E55$     &       \\  
											& $6500-6800$ & $6660$     & $0.7 \pm 0.1$  & $2.4 \pm 0.3$ & $4.4 \pm 1.0\ E55$      &    \\									
\tableline
Ridge 2 (sum) \tablenotemark{d}  	& &  						   & $3.7 \pm 0.4$  & $12.5 \pm 1.5$ & $4.3 \pm 1.1\ E56$     &   $ 0.34 \pm 0.04 $  \\
\tableline
Ridge 3                            & $5500-6100$ & $5788$    & $3.5 \pm 0.4$      & $11.6 \pm 1.7 $ & $4.5 \pm 0.6\ E56$      &   $ 0.10 \pm 0.02  $ \\							  
\tableline
Bridge                              & $5900-7000$ & $6435$    & $4.9 \pm 0.4$      & $16.0 \pm 2.0 $ & $2.4 \pm 0.5 \ E57$      &   $ 0.57 \pm 0.09  $ 
\enddata
\tablecomments{Idem as Table~\ref{table_CO10_results} but for integration of the CO(1-0) spectra over velocity ranges. The first three columns list the velocity range, the first moment velocity (mean, intensity-weighted velocity), and CO(1-0) intensity (cf. Eq.~\ref{eq:Fco_int}). Results are given for each velocity ranges, as well as for the sum of the integrations.}
\tablenotetext{a}{$\rm H_2$ gas masses calculated using the Galactic conversion factor of $N({\rm H\  _{2}})/ I_{\rm CO} = 2 \times 10^{20}$~cm$^{-2}$ [K~km~s$^{-1}$]$^{-1}$}
\tablenotetext{b}{Bulk kinetic energy, calculated from Eq.~\ref{eq:Ekin_int} (see Sect.~\ref{subsec:kin_energy}).}
\tablenotetext{c}{Ratio between the $\rm H\; _2$ mass derived from the \textit{Spitzer} spectral mapping of the mid-IR $\rm H\; _2$ lines in the group to the H$\; _2$ mass derived from the CO(1-0) line intensity at the different positions in the IGM of SQ.  }
\tablenotetext{d}{Sum over all velocity ranges at a given position.}
\end{deluxetable*}

\begin{deluxetable}{lccc}
 \tablecolumns{4}
 \tablewidth{\columnwidth}
 \tablecaption{CO(2-1) observations with the IRAM 30m telescope: results from Gaussian decompositions\label{table_CO21_results}}
\tablehead{
\multirow{2}*{Target region}   & \colhead{ $\rm v_{\rm CO(2-1)}$}   & \colhead{ $\Delta \rm v_{\rm CO(2-1)}$}   &	\colhead{ $\mathcal{I}_{\rm CO(2-1)}$}   \\	
									 &\colhead{[km s$^{-1}$]}  &  \colhead{  [km s$^{-1}$]}  &	 \colhead{  [K km s$^{-1}$]}   
}									 
\startdata
 \multirow{3}*{SQ-A}			   & $5966 \pm 25$   & $81 \pm 8$     &	$0.76 \pm 0.11$  \\
	  										       & $6719 \pm 36$   & $292 \pm 35$ &	$2.3 \pm 0.4$  \\
	  										       & $6906 \pm 14$   & $62 \pm 12$   &	$1.4 \pm 0.2$  \\
\tableline
SQ-A (tot)                                 &                              &                        &	$4.5 \pm 0.3$ \\ 
\tableline
 \multirow{4}*{Ridge~1}        & $5783 \pm 14$    & $80 \pm 10$   & $0.50 \pm 0.08$    \\
 											     & $6041 \pm 16$    & $92 \pm 12$   & $1.9 \pm 0.3$   \\
  											     & $6101 \pm 39 $   & $367 \pm 34$ & $2.4 \pm 0.4$    \\
  											     & $6721 \pm 12$    & $105 \pm 13$ & $0.74 \pm 0.12$  \\
\tableline
Ridge~1 (tot)   				            &                              &                         & $5.5 \pm 0.5$   \\
\tableline
Ridge 2 					                & $5972 \pm 28$    & $252 \pm 36$  & $2.5 \pm 0.5$  \\
\tableline  
	Ridge 3                              & $5743 \pm 25$   & $117 \pm 19$   & 	$1.9 \pm 0.4$      \\							  
\tableline
\multirow{3}*{Bridge}         & $6318 \pm 22$   & $77 \pm 12$     & 	$0.7 \pm 0.1$        \\ 
										      & $6552 \pm 46$   & $648 \pm 36$   & $5.2 \pm 1.6$          \\	
										      & $6606 \pm 24$   & $91 \pm 11$     & $0.8 \pm 0.1$          \\	
\tableline
Bridge (tot) 	& &  									                                         &  $6.7 \pm 1.6$      
\enddata
\tablecomments{Idem as Table~\ref{table_CO10_results} but for the IRAM 30m EMIR CO(2-1) observations.  The size of the CO(2-1) beam is $1.13 \times (11\ \rm arcsec)\ ^2 = 137 \  \rm arcsec\,^2$. }
\end{deluxetable}

\begin{deluxetable}{lccc}
 \tablecolumns{4}
 \tablewidth{\columnwidth}
 \tablecaption{CO(3-2) observations with APEX: results\label{table_CO32_results}}
\tablehead{
\colhead{ \multirow{2}*{Target}}   & \colhead{ $\rm v_{\rm CO(3-2)}$}   & \colhead{ $\Delta \rm v_{\rm CO(3-2)}$}   &	\colhead{ $\mathcal{I}_{\rm CO(3-2)}$}     \\	
									 &\colhead{  [km s$^{-1}$]}  &  \colhead{  [km s$^{-1}$]}  &	 \colhead{  [K km s$^{-1}$]}    
}									 
\startdata
   SQ-A  \tablenotemark{a} 		   & $6858 \pm 66$   & $80 \pm 31$ &	$0.32 \pm 0.12$    \\
\tableline
Ridge~1         & $5976 \pm 15$    & $134 \pm 43$      & $0.65 \pm 0.13$  
\enddata
\tablecomments{Idem as Table~\ref{table_CO10_results} and \ref{table_CO21_results} but for  the APEX CO(3-2) observations (see Figure~\ref{fig:SQ_CO32_spectra}) obtained with the FLASH345/XFFTS receivers.  The aperture of the CO(3-2) beam is $1.13 \times (18.1\ \rm arcsec)\ ^2 \approx 370 \ \rm arcsec\,^2$. }
\tablenotetext{a}{Tentative detection at a 2.3~$\sigma$ level.}
\end{deluxetable}

\begin{figure}
   \centering
   \includegraphics[width=\columnwidth]{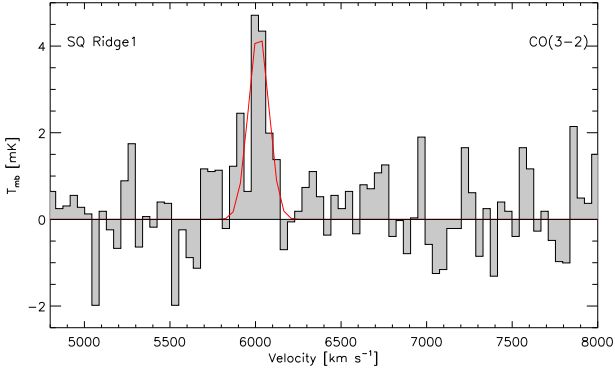}
      \caption{
      CO(3-2) spectra observed with APEX at the position Ridge~1  in the IGM of the SQ group (see Figure~\ref{fig:CO_beams_on_HST}).  The red line shows the result of the line fitting (see Tables~\ref{table_CO32_results}). The spectrum has been smoothed to a spectral resolution of 40~km~s$^{-1}$.
      }
       \label{fig:SQ_CO32_spectra}
   \end{figure}

\begin{figure}
   \centering
    \includegraphics[width=\columnwidth]{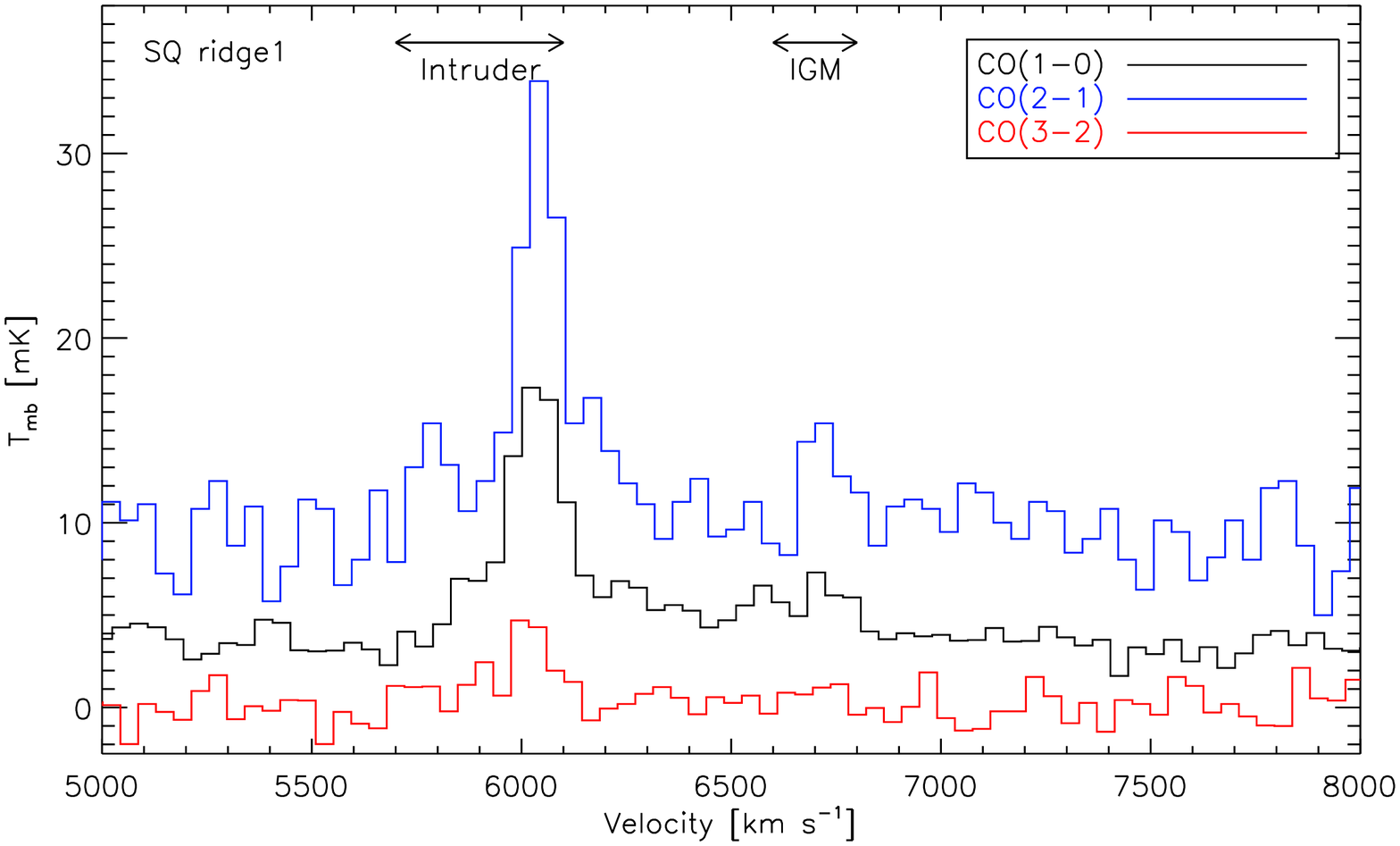}
      \caption{Overlay of the CO(1-0), CO(2-1) and CO(3-2) spectra observed in the Stephan's Quintet shock, at the position Ridge~1. No beam size convolution has been applied. The spectra have been shifted vertically for clarity. The arrows indicate the pre-shock H{\sc i} velocities for the intruder galaxy and the intra-group medium.}
       \label{fig:SQ_CO32_21_10_spectra}
   \end{figure}

\subsection{CO(3-2)  with the APEX telescope}
\label{subsec:CO_obs_apex}

Observations towards the Stephan's Quintet in the CO(3-2) transition were carried out during night time in July 2010 with the Atacama Pathfinder EXperiment telescope \citep[APEX, ][]{Gusten2006}, using the First Light APEX Submillimeter Heterodyne receiver (developed by Max Planck Institut f\"{u}r Radioastronomie, MPIfR, and commissioned in May 2010) at 345~GHz, FLASH345\footnote{In the 345 GHz band the dual-polarization receiver FLASH operates a 2SB SIS mixer provided by IRAM \citep{Maier2005}.}, along with with two newly commissioned MPIfR X-Fast Fourier Transform Spectrometer backends, XFFTS (Klein et al., in preparation), of 2.5~GHz bandpass each, and whose combination covers a 4~GHz bandpass. 
The beam of the telescope is 18.1" at 345~GHz \citep{Gusten2006}. Focus was checked at the beginning of each night, and pointing was checked on planets (Jupiter, Saturn, and/or Uranus), as well as on CRL2688 every 90 minutes at least.

System temperatures were in the range 180-270~K and 180-230~K in the respective cases of the Ridge~1 and SQ-A positions, on the T$_{A}^{\*}$ scale. Antenna temperatures were converted to main beam temperatures using forward and beam efficiencies, respectively, of 0.95 and 0.67 at 345~GHz. We used 32768 channels for each XFFTS backends, yielding an effective spectral resolution of about 76.3~kHz. The observations were performed in wobbler switching mode, with a throw of 120" in the azimuthal direction. 

Again, the data was reduced by means of the IRAM GILDAS software package. We noted that $\approx 15$ channels (out of a total of 32768 covering the $4800-8400$~km~s$^{-1}$ velocity range) were affected by spurious spikes, likely due to the electronics of other receivers that were powered-on at the time of the observations). Therefore, a sigma-clipping has been applied. By scanning across all spectra, the sigma associated with each channel, $\sigma _c$, is determined. The data is flagged if the signal is greater than $5\,\sigma _c$ and greater than $3.5 \, \sigma _s$, where $\sigma _s$ is the sigma associated with one spectrum, since the signal does not reach this level.
																															
The baseline subtraction was always linear, as we enjoyed very good baseline stability due to excellent weather conditions during the observations. The final spectra shown here are smoothed to a velocity resolution of 40~km~s$^{-1}$, for comparisons purposes with the other transitions obtained at the IRAM 30m-telescope. 

\section{Kinematics and excitation of the CO gas}
\label{sec:kinetics_CO}

\subsection{CO spectra and analysis}

In Figure~\ref{fig:SQ_CO_spectra} we show the CO(1-0) and CO(2-1) spectra for each of the five observed positions. 
The spectra are plotted over a $4000$~km~s$^{-1}$ bandwidth (out of the $\approx 8000$~km~s$^{-1}$ bandwidth available at 115~GHz) at a spectral resolution of 40~km~s$^{-1}$. The noise levels in the average spectra are typically $\approx 0.7 - 1.6$~mK for CO(1-0), and $\approx 2 - 5$~mK for CO(2-1) at this resolution.  The two transitions of CO are detected with a high signal-to-noise ratio for all the targeted regions, except a $2.5\,\sigma$ tentative detection of CO(2-1) for the R3 position. Such detections are in sharp contrast with the previous interferometric observations, where no CO was detected in the ridge, probably because of limited velocity coverage and sensitivity. It shows that CO gas is present in the H$_2$-luminous shock, and is co-spatial with the X-ray emitting hot plasma in this region, at a $\sim 5\,$kpc scale.

The spectrum resulting from our pilot study with APEX at the ridge~1 position, where the CO(3-2) line is detected, is shown in Figure~\ref{fig:SQ_CO32_spectra}. In SQ-A, the line is tentatively detected at a 2.3$\,\sigma$ significance, and we consider this as an upper limit on the CO(3-2) flux.

To compute the CO line parameters, we applied two different methods. First, we decomposed the CO spectra in multiple Gaussian components using the IDL non-linear least-squares MPFIT routine \citep{Markwardt2009}. The red lines on Figures~\ref{fig:SQ_CO_spectra} and \ref{fig:SQ_CO32_spectra} show the fitting result, which is used to derive the central velocities of the CO components,  the line widths and integrated intensities $\mathcal{I}_{\rm CO}$. We checked that the residuals are below the 2$\, \sigma$ noise level for all positions and all lines.  The results of these Gaussian decompositions are given in Tables~\ref{table_CO10_results}, \ref{table_CO21_results} and \ref{table_CO32_results}.

Second, the CO line intensities $\mathcal{I}_{\rm CO}$ were derived from the integration of the CO main beam temperature, $\mathcal{T}_{\rm mb}$~[K], over velocity ranges, from $\rm v_{min}$ to $\rm v_{max}$, defined for each spectrum in Table~\ref{table_CO10_results_int}, as the following:
\begin{equation}
\label{eq:Fco_int}
\mathcal{I}_{\rm CO} \ {\rm [K\ km\ s^{-1}]} = \int _{\rm  \langle \rm v \rangle - v_{min}} ^{\rm v_{max} -  \langle \rm v \rangle} \mathcal{T}_{\rm mb} \; d\rm v \ ,
\end{equation}
where $ \langle \rm v \rangle$ is the first moment (mean, intensity-weighted) velocity over that velocity range.
This method allows us to better take into account the asymmetry in the CO profiles.  The results of these integrations are given in Table~\ref{table_CO10_results_int} and discussed in the following sections.

\subsection{Multiple CO velocity components}
\label{subsec:multiple_vel}

The spectra in Figures~\ref{fig:SQ_CO_spectra} and \ref{fig:SQ_CO32_spectra} clearly show that multiple, broad velocity components are detected, pointing out the complexity of the  kinematics of the CO gas in the SQ group. The CO line emission extends over a very wide velocity range $\approx 5700 - 7000$~km~s$^{-1}$ (and perhaps until $7400$~km~s$^{-1}$ for the ridge~1 position).  The four main velocity components are at 5700, 6000, 6700 and 6900~km~s$^{-1}$ (see table~\ref{table_CO10_results}, \ref{table_CO21_results} and \ref{table_CO32_results} for the central line velocities derived from line fitting). The gas associated with the intruder NGC~7318b ($5700-6000$~km~s$^{-1}$) and with the intra-group medium ($6700$~km~s$^{-1}$) are both detected in the ridge and in the eastern feature (the bridge) towards NGC~7319. 
These two CO velocity systems match with the velocities of the H{\sc i} gas observed in the group \citep{Sulentic2001, Williams2002}.   
In Ridge~3, the southern part of the shock, most of the gas seems to be associated with the intruder, since the $6700$~km~s$^{-1}$ component is not detected. The velocity component at 6900~km~s$^{-1}$ is a new feature that we detect only in the SQ-A region. It was not seen in previous CO or H$\,${\sc i} observations because of limited velocity coverage, and its origin is still unknown. It could be the result of  dynamical effects during the previous encounter between NGC 7319 and 7320c, but current hydrodynamical simulations cannot reproduce this feature. 
We note that an H$\alpha$ velocity component at 7000~km~s$^{-1}$ was reported by \citet{Sulentic2001} in  a clump to the North-East of SQ-A, but it is outside of our CO(1-0) beam. 

Gas at intermediate velocities, in between that of the intruder and that of the intra-group, is detected. A weak, broad component centered at $\approx 6400$~km~s$^{-1}$ is detected in the ridge~1, ridge~2 and in the bridge positions. Interestingly, this intermediate component seems to be absent (or very weak, at a $\approx 2\sigma$ significance) in SQ-A and vanishes at the southern end of the main shock region (ridge~3). The $6400$~km~s$^{-1}$ CO velocity component is consistent with the central velocity of the broad, resolved $\rm H_2$ S(1) line at  $6360 \pm 100$~km~s$^{-1}$ \citep[see ][for an analysis of the high-resolution \textit{Spitzer} IRS spectrum]{Appleton2006, Cluver2010}. 
This is the first time CO gas is detected at these intermediate velocities, and this gas is spatially associated with the warm $\rm H_2$ seen by \textit{Spitzer}. 
\citet{Xu2003, Iglesias-Paramo2012} reported that the low velocity component (intruder) of the optical emission lines in the shock and in the bridge is centered at intermediate velocities (between 5900 and 6500~km~s$^{-1}$). In the northern (SQ-A) and southern star-forming regions, at both ends of the large-scale shock, the low velocity component of the optical lines is consistent with the intruder velocity. 

In Figure~\ref{fig:SQ_CO32_21_10_spectra} we compare the three CO line profiles at the ridge~1 and SQ-A positions. At both positions, the central velocity of the CO(3-2) line coincide with the brightest CO(1-0) and (2-1) components. The FWHM of the CO(3-2) line matches those of the brighter CO(1-0) and (2-1) lines. 

\subsection{High CO velocity dispersions}
\label{subsec:high_CO_vel_disp}

In the \textit{ridge}, the FWHM of the main CO components are of the order of $100-400$~km~s$^{-1}$ (see tables~\ref{table_CO10_results}, \ref{table_CO21_results} and \ref{table_CO32_results}). In the ridge~1 position (and perhaps ridge~2), the underlying intermediate component at $6400$~km~s$^{-1}$ is extremely broad, $\approx 1000$~km~s$^{-1}$, with possibly a hint of emission between 6900 and 7400~km~s$^{-1}$.  This CO velocity coverage of  $\approx 1000$~km~s$^{-1}$ is comparable to the relative velocity between the intruder and the IGM, and to the broadening of the H$_2$ S(1) line (FWHM$\approx 870$~km~s$^{-1}$) observed by \citet{Appleton2006, Cluver2010}. 

In \textit{SQ-A}, the linewidths are of the order of 100~km~s$^{-1}$, with the smallest dispersion for the 6000~km~s$^{-1}$ line. The broadest component at 6700~km~s$^{-1}$ seems to have a left wing extending towards 6400~km~s$^{-1}$, suggesting again that some of the CO gas belonging to the IGM has been decelerated in the shock induced by the intruder galaxy.
Interestingly, for the three transitions, the CO linewidths are significantly smaller in SQ-A than in the ridge and bridge, and this is perhaps a clue to understand why there is much more star formation in this region than in the ridge (see discussion in sect.~\ref{subsec:shock_SF}). This observational picture fits within the interpretation of two colliding gas flows, where the shear velocities between the flows are maximum in the central region of the contact discontinuity (ridge), and smaller at the edges of the main shock structure (see Figure~5 of \citet{Lee1996} and \citet{GuillardP.2010}). 

The spectrum in the \textit{bridge} is perhaps the most striking. It shows a very broad signal in between the extreme velocities detected in SQ-A (6000 and 6900~km~s$^{-1}$), see Figure~\ref{fig:SQ_CO_spectra}.  The CO(1-0) profile is well fitted by a sum of 2 Gaussians, whereas in the CO(2-1) spectrum, narrower velocity components (centered at 6300 and 6600~km~s$^{-1}$) emerge on top of a broader (FWHM$ \approx 650$~km~s$^{-1}$) emission.

\subsection{CO excitation}

Since the beams of the CO(1-0), (2-1) and (3-2) observations do not match, and given that the angular size of the CO emitting regions are not constrained, we cannot infer useful information on the CO excitation (from the CO line ratios). 
We generally find a CO(2-1) to CO(1-0) line ratio of $\approx 1$, which suggests that the CO is not far from being optically thick. This has to be confirmed with  a complete CO(2-1) mapping of  SQ.

\section{Distribution and mass of molecular gas}
\label{sec:distribution_mass_CO}

\subsection{Warm H$_2$ masses from the \textit{Spitzer} mid-IR spectroscopy}

To derive the warm ($T \gtrsim 100$~K) H$_2$ masses from the \textit{Spitzer} observations, we extracted IRS spectra within the areas observed in CO(1-0) with the CUBISM software \citep{Smith2007a}. The H$_2$ line fluxes, listed in Table~\ref{table_H2line_intensities}, were derived from Gaussian fitting with the PAHFIT IDL tool \citep{Smith2007}. The H$_2$ physical parameters are estimated with the method described in \citet{Guillard2009}. The  spectral energy distribution of the H$_2$ line emission is modelled with magnetic shocks, using the MHD code described in \citet{Flower2010}. The gas is heated to a range of post-shock temperatures that depend on the shock velocity, the pre-shock density, and the intensity of the transverse magnetic field. We use a grid of shock models (varying shock speeds) similar to that described in \citet{Guillard2009}, to constrain the physical conditions (density, shock velocity) needed to fit the H$_2$ line fluxes. These fits are not unique, and in principle, the observed H$_2$ excitation arises from a continuous distribution of densities and shock velocities. We find a best fit for a pre-shock density $n_{\rm H} = 10^3$~cm$^{-3}$. The initial ortho-to-para ratio is set to 3, and the intensity of the pre-shock magnetic field is set to $B_0 = \sqrt{n_{\rm H}} \approx 30\,\mu$G, in agreement with Zeeman effect observations of  Galactic molecular clouds \citep{Crutcher1999}, and the value inferred from radio observations of the synchrotron emission in the SQ ridge \citep{Xu2003, O'Sullivan2009a}. The H$_2$ line fluxes and the warm H$_2$ masses are computed when the post-shock gas has cooled down to a temperature of 100~K. This temperature is chosen because below $\approx 100$~K, the gas does not contribute significantly to the rotational H$_2$ emission. The warm H$_2$ masses derived from the shock models depend on this temperature, the lower the temperature, the longer the cooling time, hence the larger H$_2$ mass. However, the relative difference in cooling times (hence in H$_2$ masses) between 100~K and 150~K is $< 20\,$\%, so this choice of temperature does not impact much the results given in  Table~\ref{table_shock_param}.

\begin{deluxetable*}{l c c c c c c}
 \tablecaption{$\rm H_2$ rotational line intensities}
 \tablecolumns{7}
 \tablewidth{0pt}
\tablehead{
\colhead{Target region} & \colhead{$\rm H_2$ S(0)} & \colhead{$\rm H_2$ S(1)} & \colhead{$\rm H_2$ S(2)} & \colhead{$\rm H_2$ S(3)} & \colhead{$\rm H_2$ S(4)} & \colhead{$\rm H_2$ S(5)} }
\startdata
SQ-A 	  	& $ 1.47 \pm 0.11 $ & $ 6.26 \pm 0.13 $ &   \\		
ridge~1   	& $ 0.79 \pm 0.04 $ & $ 11.0 \pm 1.56 $ & $ 3.19 \pm 0.25 $ & $ 11.5 \pm 1.31 $ & $ 1.97 \pm 0.27 $ & $ 4.97 \pm 0.41 $ \\
Ridge 2  	& $ 0.68 \pm 0.04 $ & $ 7.06 \pm 0.42 $ & $ 2.75 \pm 0.11 $ & $ 13.3 \pm 2.35 $ & $ 1.78 \pm  0.33$ & $ 4.31 \pm 0.31 $ \\
Ridge 3   	& $ 0.49 \pm 0.07 $ & $ 9.55 \pm 1.51 $ & $ 3.42 \pm 0.31 $ & $ 5.52 \pm 1.53 $ & $ 0.94 \pm 0.34 $ & $ 2.85 \pm 0.39 $ \\
Bridge	  	& $ 0.88 \pm 0.07 $ & $ 4.88 \pm 0.08 $ &   
\enddata
\tablecomments{$\rm H_2$ line fluxes derived from the \textit{Spitzer IRS} spectra extracted at the positions of our CO observations. Fluxes are in units of $10^{-17}$ W~m$^{-2}$.}
\label{table_H2line_intensities}
\end{deluxetable*}

\begin{deluxetable}{lcccc} 
 \tablecaption{$\rm H_2$ line flux fitting with MHD shock models: results}
\tablecolumns{5}
\tablewidth{\columnwidth} 
\tablehead{
\colhead{ \multirow{2}*{Target region}} 	& \colhead{shock velocities}    & \colhead{Mass flow} 			  & \colhead{Cooling time} 		  & \colhead{$M(\rm H_2^{\rm warm})$}   \\
													& \colhead{[km~s$^{-1}$]}    & \colhead{[M$_{\odot}$~yr$^{-1}$]} & \colhead{[yr]}		  & \colhead{[M$_{\odot}$]} }
\startdata
\tableline
SQ-A 	  										& 		5					 & $1.47\times 10^4$& $8.09\times 10^4$& $1.19\times 10^9$	 \\	
\tableline	
 \multirow{2}*{Ridge~1}  			& 8							 & $6.28\times 10^3$& $5.33\times 10^4$& $3.35\times 10^8$  \\
 													& 35    					 & $5.45\times 10^2$& $9.77\times 10^3$& $5.33\times 10^6$	 \\	
\tableline
 \multirow{2}*{Ridge 2}  			& 6							 & $6.61\times 10^3$& $6.30\times 10^4$ & $4.16\times 10^8$  \\
 													& 35    					 & $5.40\times 10^2$& $9.77\times 10^3$ & $5.27\times 10^6$	 \\	 
 \tableline
 \multirow{2}*{Ridge 3}   			& 13					     & $3.32\times 10^3$& $5.52\times 10^4$& $1.17\times 10^8$  \\
  													& 40     					 & $1.74\times 10^2$& $7.76\times 10^3$& $1.35\times 10^6$	 \\	
\tableline  													
Bridge	  										& 5							 & $1.13\times 10^4$& $8.09\times 10^4$& $9.13\times 10^8$	 	
\enddata 
\tablecomments{$\rm H_2$ line fluxes are fitted with low-velocity MHD shocks in dense molecular gas, for a pre-shock density of $n_{\rm H} = 10^3\,$cm$^{-3}$ and a pre-shock magnetic field of $30\,\mu$G. The table gives the best-fit shock velocities, mass flows, cooling times and warm $\rm H_2$ masses.}
\label{table_shock_param}
\end{deluxetable}

A combination of two shocks with distinct velocities is required to match the observed H$_2$ line fluxes.  As discussed in \citet{Guillard2009}, the H$_2$ masses are derived by multiplying the gas cooling time (down to 100~K) by the gas mass flow (the mass of gas swept by the shock per unit time) required to match the H$_2$ line fluxes. The shock model parameters, gas cooling times, mass flows, and warm H$_2$ masses are quoted in Table~\ref{table_shock_param}. The IRS spectra in different regions of the shock, and a detailed description of the spatial variations of H$_2$ excitation (full 2-dimension excitation diagrams) in the SQ intra-group medium will be given in a separate paper (Appleton et al., in preparation).

\subsection{H$_2$ masses derived from CO observations}
\label{subsec:CO_mass_distrib}

We derive the masses of H$_2$ gas from the CO(1-0) line observations, using the Galactic conversion factor 
$X = N(\rm H_2) / \mathcal{I}_{\rm CO} = 2 \times  10^{20}~\rm cm^{-2}~(K~km~s^{-1})^{-1}$:
\begin{eqnarray}
\label{eq:coldH2mass}
M_{\rm H_2} & \!=\! &  \mathcal{I}_{\rm CO} ^{\rm obs} \times \frac{N(\rm H_2) }{\mathcal{I}_{\rm CO}} \times d^2 \times \Omega \times 2 \, m_{\rm H} \\
\nonumber
\frac{M_{\rm H_2}}{\rm M_{\odot}}	& \!= \!& 3.3 \!\times \!10^8  \frac{ \mathcal{I}_{\rm CO} ^{\rm obs}}{\rm K \, km \, s^{-1}} \frac{N(\rm H_2) / \mathcal{I}_{\rm CO}}{2 \! \times \! 10^{20}} \left(\frac{d}{94 \, \rm Mpc} \right)^2 \left(\frac{\theta _{\rm {\tiny HP}}}{22''} \right) ^2 
\end{eqnarray}
where $ \mathcal{I}_{\rm CO} ^{\rm obs}$ is the observed velocity integrated CO(1-0) line intensity, $d$ is the distance to the source, and $\Omega$ is the area covered by the observations in arcsec$^2$, with $\Omega = 1.13 \times \theta _{\rm \tiny HP} ^{2}$ for a single
pointing with a Gaussian beam of Half Power beam width of $\theta _{\rm \tiny HP} ^{2}$. To derive the total mass of molecular gas, one has to multiply $M_{\rm H_2}$ by a factor 1.36 to take the Helium contribution into account. Tables~\ref{table_CO10_results} and \ref{table_CO10_results_int} give the  $\rm H_2$ gas masses for the different positions, derived from the CO line intensities. 
In the following, we discuss the masses of molecular gas detected in the three main areas, the ridge, SQ-A, and the bridge.

In the SQ \textit{ridge} (positions R1, R2, and R3), summing over all the CO(1-0) velocity components, and assuming a Galactic CO to H$_2$ conversion factor, we derive a mass of $\approx 4.1 \times 10^9$~M$_{\odot}$ of  $\rm H_2$ gas, corresponding to a surface density of  $\Sigma(\rm H_2) = 12.0 \pm 1.3$~M$_{\odot}$~pc$^{-2}$.  Note that these three pointings cover $\approx 75$\% of the whole H$_2$-bright ridge (the R1 and R2 beams are partially overlapping). 

Within the CO(1-0) aperture  $\Omega = 1.13 \times (22 \ \rm arcsec)^2 = 547$~arcsec$^2$ centered on \textit{SQ-A}, we find a total  $\rm H_2$ mass of $1.3 \times 10^9$~M$_{\odot}$. \citet{Lisenfeld2002} find $\approx 3 \times 10^9$~M$_{\odot}$ over a much larger area ($60 \times 80 \ \rm arcsec ^2$) that partially overlaps the north of the ridge.
However, because of a limited bandwith, they only detected the 6000 and 6700~km~s$^{-1}$ velocity components, and not the higher velocity component at 6900~km~s$^{-1}$ that  contains almost half of the mass in the core region of SQ-A (see Table~\ref{table_CO10_results}). 
Based on the MIPS $24\,\mu$m map, we estimate the area of SQ-A to be $\approx 40 \times 40 \ \rm arcsec ^2$. If we scale the mass we find in our aperture $\Omega$ to this larger aperture, we find a total mass of  $\rm H_2$ of $3.8 \times 10^9$~M$_{\odot}$. This is comparable to the mass derived by \citet{Lisenfeld2002}, because the larger aperture used to estimate their masses compensates their non-detection of the 6900~km~s$^{-1}$  component.

In NGC~7319's \textit{bridge}, summing over all velocities, we find $M_{\rm H_2} \approx 1.7 \times 10^9$~M$_{\odot}$ within the aperture $\Omega$. Note that this aperture  matches quite well the width of the bridge, but does not cover its whole extension towards the AGN. Therefore, the total mass of molecular gas in the bridge may be a factor of $\approx 2$ larger. \citet{Smith2001a} find $\approx 5 \times 10^9$~M$_{\odot}$ within a half-power beam of 55~arcsec in diameter centered on the nucleus of NGC~7319. This beam overlaps our bridge region.
The global morphology of the CO gas in SQ is still unknown, but it is surely linked to the geometry of the collision and the morphology of the colliding gas flows. For instance, the origin of the molecular gas in the bridge structure is still an open question. 
 It is most likely a result of a previous tidal interaction (with NGC~7320c for instance) that would have stripped some material from NGC~7319's galactic disk. Then the interaction between the new intruder and this ``tidal bridge'' may trigger further molecular gas formation in this region, since the proportion of post-shock gas (with respect to the pre-shock) in this region is the highest observed. \citet{Aoki1996} also reported from optical spectroscopy that an outflow from the AGN  is present, which seems oriented in the direction of the bridge feature.  Some of the molecular material present in the bridge may be gas entrained from NGC~7319's disk by the outflow.  However,  it is unlikely that the outflowing ionized gas can entrain the dense molecular gas on such large distances ($\gtrsim 10$~kpc).

These observations represent a substantial revision of the mass and energy budgets of the collision. Previous interferometric observations may have missed the molecular gas in the ridge because of  limited sensitivity and bandpass and because the CO emission in the ridge may be more diffuse than in SQ-A.  We detect at least 4 times more CO emission in the SQ ridge and in the bridge than in the starburst region SQ-A. Although these observations do not cover all the IGM, the CO emission seems distributed over a 40~kpc scale along the ridge, and in total (SQ-A + ridge + bridge), $\gtrsim 5 \times 10^9$~M$_{\odot}$ of $\rm H_2$ gas is lying outside the optical disks of the group galaxies. Note that these masses depend on the $X$-factor, which is poorly constrained (sect~\ref{subsec:CO_mass_errors}).

\subsection{Uncertainties in the H$_2$ masses}
\label{subsec:CO_mass_errors}

We computed the uncertainties in the warm H$_2$ masses derived from the \textit{Spitzer} spectroscopy by fitting shock models to the lower and upper values of the H$_2$ line fluxes, and we found uncertainties on the H$_2$ masses of the order of 20\% (based solely on the observational uncertainties). The warm H$_2$ masses derived from shock models  are $10-35\,$\% larger than local thermal equilibrium LTE models (we performed single temperature fits for SQ-A and the Bridge regions, and two temperatures fits for the positions in the ridge). This is mainly because (i) the values of the H$_2$ ortho-to-para ratio are different in the two models, (ii) at $n_{\rm H} = 10^3$~cm$^{-3}$, the S(0) and S(1) lines are not fully thermalized, and (iii) the H$_2$ mass derived from shock models is dependent on the final temperature at which the H$_2$ level populations are calculated.

We derived H$_2$ masses from the CO(1-0) line intensities using the Galactic value of the CO-to-H$_2$ mass conversion factor (the $X$-factor), because the metallicity in the IGM is slightly above the solar value \citep{Xu2003}. However, 
the $X$-factor is uncertain, essentially because it  is poorly constrained in shocked environments  \citep{Downes1998, Gao2003}, and more generally in low density molecular gas. 
 For instance, in a similar situation in the bridge between the Taffy Galaxies (UGC~12914 and 12915), \citet{Braine2003} estimated a $X$-factor roughly five times below the Galactic value using $^{13}$CO measurements. 
The dust emission modelling by \citet{Natale2010}, using their \textit{Spitzer}/MIPS 70 and 160$\,\mu$m measurements in the shock, gives $\Sigma(\rm H_2) = 9.8$~M$_{\odot}$~pc$^{-2}$ (assuming a Galactic dust-to-gas mass ratio $Z_{\rm d} = 0.0075$).
This value is close ($\approx 20\,$\% smaller) to the average H$_2$ mass surface  density derived from the CO measurements in the ridge ($\Sigma(\rm H_2) = 12.0 \pm 1.3$~M$_{\odot}$~pc$^{-2}$), assuming a Galactic conversion factor (Table~\ref{table_CO10_results}). This is within the uncertainties of the derived values from the dust modeling. The large uncertainties on the far-IR photometry in the shock are  dominated by the difficulty to spatially separate the far-IR emission of the ridge itself from the bright sources surrounding the shock. 
Upcoming \textit{Herschel} observations will help in estimating the $X$-factor in the shock and in the surrounding star forming regions. We defer this discussion to a future paper. 
The uncertainty on the $X$-factor affects the mass and energy budget discussed in Sect.~\ref{subsec:mass_nrj_budgets} but does not change  the general conclusions of this paper.

\section{Energetics and excitation of the molecular gas}
\label{sec:excit_nrj}

\subsection{Kinetic energy carried by the molecular gas}
\label{subsec:kin_energy}

A first estimate of the kinetic energy of the molecular gas was based on \textit{Spitzer}/IRS spectroscopy of the H$_2$ line emission from SQ \citep{Guillard2009}.  Assuming that the $8.8\times 10^8$~M$_{\odot}$ of warm H$_2$ gas detected by \textit{Spitzer} has a velocity dispersion of $370$~km~s$^{-1}$ \citep{Cluver2010}, the turbulent kinetic energy carried by the warm H$_2$ gas is $\approx 3.6\times 10^{57}$~ergs. Given the limited spectral resolution of \textit{Spitzer}/IRS ($\approx 600$~km~s$^{-1}$), this estimate was crude. The present CO observations allows us to compute the kinetic energy more accurately.

As for the CO fluxes and masses, we estimate the kinetic energy carried in the CO emitting gas with two methods. First, from the Gaussian decompositions, the turbulent kinetic energy of the H$_2$ gas is derived from $E_{\rm kin} = 3/2 M_{\rm H_2} \sigma _{\rm CO}^{2}$, where $M_{\rm H_2}$ is the H$_2$ mass derived from the CO(1-0) flux  (Eq.~\ref{eq:coldH2mass}) and $\sigma _{\rm CO}$ the velocity dispersion along the line of sight (measured on the Gaussian fits of the CO(1-0) line). The factor of 3 in $E_{\rm kin}$ takes into account the three dimensions. These kinetic energies are listed in Table~\ref{table_CO10_results} for each of the CO velocity component and each observed positions.

The second method consists in integrating $\mathcal{T}_{mb} \times (\rm v-\langle v \rangle)^2$ over velocity ranges, from $\rm v_{min}$ to $\rm v_{max}$, defined for each spectrum and each velocity components in Table~\ref{table_CO10_results_int}:
\begin{eqnarray}
\label{eq:Ekin_int}
 E_{\rm kin} & = & \frac{3}{2} \times 3.3 \times 10^8 \rm M_{\odot} \frac{N(\rm H_2) / \mathcal{I}_{\rm CO}}{2 \! \times \! 10^{20}}  \left(\frac{d}{94 \, \rm Mpc} \right)^2 \left(\frac{\theta _{\rm {\tiny HP}}}{22''} \right) ^2 \nonumber \\ 
                      & & \times \int _{\langle v \rangle - \rm v_{min}} ^{\rm v_{max}-\langle v \rangle} (\rm v-\langle v \rangle)^2 \times \mathcal{T}_{mb} \, d\rm v \ ,
 \end{eqnarray}
where $\langle\rm  v \rangle$ is the mean, intensity-weighted (first moment) velocity. The comparison of Tables~\ref{table_CO10_results} and \ref{table_CO10_results_int} shows that the two methods give similar results, essentially because we defined velocity ranges that bracket the main peaks of CO emission. Integrations over these velocity ranges give lower limits of the bulk kinetic energy of the H$_2$ gas for each positions. An integration over the full extent of the CO velocity range gives an upper limit (see Table~\ref{table_CO10_results_int}).

The CO emitting gas carries a huge amount of  kinetic energy, $\approx 4\times 10^{57}$~ergs when summing on all positions. 
The kinetic energy of the H$_2$ gas is at least 5 times higher than the thermal energy of the hot plasma heated by the galaxy collision \citep[][estimated that this thermal energy is $\approx 9 \times 10^{56}\,$ergs for the entire ridge and bridge]{Guillard2009, Guillard2010}. 
Before the collision, much of the kinetic energy of the gas is associated with the motions of the intruder and intra-group H{\sc i} gas. The pre-shock H{\sc i} mass in the area covered by the CO observations is estimated to be $1.0-2.8 \times 10^9$~M$_{\odot}$ \citep[based on an extrapolation of the H{\sc i} observations in NGC~7319's tidal tail to the area of the shock, see sect.~2.2.1 of][]{Guillard2009}. Assuming that this gas is shocked at a velocity of 600~km~s$^{-1}$, the pre-shock H{\sc i} kinetic energy is $0.5-1.3 \times 10^{58}$~ergs. 
Therefore, the CO emitting gas is carrying a large fraction of the pre-shock kinetic energy of the gas. Most of this  kinetic energy has not been thermalized yet. We suggest that this turbulent kinetic energy is the energy reservoir that powers the H$_2$ line emission and maintains a large fraction of the mass of the H$_2$ gas warm in SQ.


\subsection{Turbulent heating of the molecular gas}
\label{subsec:warm2coldH2ratio}

To assess the physical state of the molecular gas, we first compute the ratio  between the warm H$_2$ mass derived from the \textit{Spitzer/IRS} spectroscopy and the H$_2$ mass derived from the CO(1-0) observations, which we call \textit{``mid-IR/CO H$_2$ mass ratio''}.
 These ratios are listed for each observed position in Table~\ref{table_CO10_results}.
On average, this ratio is much higher in SQ than in star-forming galaxies, where the mass of cold molecular gas ($<100$~K, usually traced by CO) is $25-250$ times greater than the warm ($> 100$~K) $\rm H_2$ mass detected in the mid-IR \citep{Roussel2007}. 
The H$_2$ mass in the ridge derived from the CO(1-0) line emission is only a factor of $\approx 5$ larger than the warm  $\rm H_2$ mass seen in the mid-IR domain ($8.8 \times 10^8 $~M$_{\odot}$, see Table~\ref{table_shock_param}).
Surprisingly, this H$_2$ mass ratio is extreme in SQ-A, whereas in the southern star-forming region (ridge~3) it is the lowest, with a value (0.1) compatible with star forming galaxies.

The mid-IR/CO H$_2$ mass ratios in SQ are  comparable to those observed in molecular clouds near the Galactic center \citep{Rodr'iguez-Fern'andez2001}, or in the bridge between the Taffy galaxies (UGC 12914/5), where $0.1 \lesssim M_{\rm H_2}^{\rm (w)} / M_{\rm H_2}^{\rm (c)}  \lesssim 0.7$ (Peterson et al., submitted), or in H$_2$-luminous radio galaxies \citep{Ogle2010}. In all these environments, it has been concluded that the molecular gas is heated by shocks rather than UV photons. 
In such environments, the CO emission may trace not only cold molecular gas ($\approx 10-100\,$K) but also warmer gas ($\approx 100$ to a few $10^3$~K, like the gas traced by mid-IR rotational H$_2$ lines).

The H$_2$ luminosity-to-mass ratio, $\mathcal{L}(\rm H_2) / \rm M (\rm H_2)$ measures the cooling rate (per unit mass) of the molecular gas \citep{Guillard2009, Nesvadba2010}. 
We compare this ratio to the turbulent heating rate per unit mass, written as $\Gamma _{\rm T} = \frac{3}{2} \sigma _{T} ^{2} / t_{\rm dis}$ \citep{MacLow1999}. $\sigma _{T}$ is the turbulent velocity dispersion of the H$_2$ gas, and $t_{\rm dis}$ is the dissipation timescale, often written as $t_{\rm dis} = l / \sigma _{T}$, where $l$ is the size of the region over which the velocity dispersion is measured. If the H$_2$ emission is powered by the dissipation of turbulent energy, then the total energy dissipation rate must be at least equal to the H$_2$ luminosity \citep{Guillard2009}:
\begin{equation}
\frac{3}{2}  \frac{\sigma _{T}^{3}}{l} \geqslant \frac{\mathcal{L}(\rm H_2)}{M(\rm H_2)}
\end{equation}
On average in the SQ IGM, we find $\mathcal{L}(\rm H_2) / \rm M (\rm H_2) = 0.10 \pm 0.02$~L$_{\odot}$~M$_{\odot}^{-1}$. At the spatial scales of a GMC ($l \approx 100\,$pc), this condition implies $\sigma _{T} > 35\,$km~s$^{-1}$. This is an order of magnitude higher than the CO velocity dispersion  observed in Galactic GMCs at these spatial scales \citep{Heyer2009}, which shows that the molecular gas in SQ is much more turbulent than in the Galaxy.

\section{Discussion}
\label{sec:discussion}

These CO observations  raise questions related to the efficiency of  star formation in the IGM and the origin of the CO gas. 

\subsection{Shocks, star formation, and physical state of the molecular gas in the SQ group}
\label{subsec:shock_SF}

\begin{figure}
   \centering
    \includegraphics[width=\columnwidth]{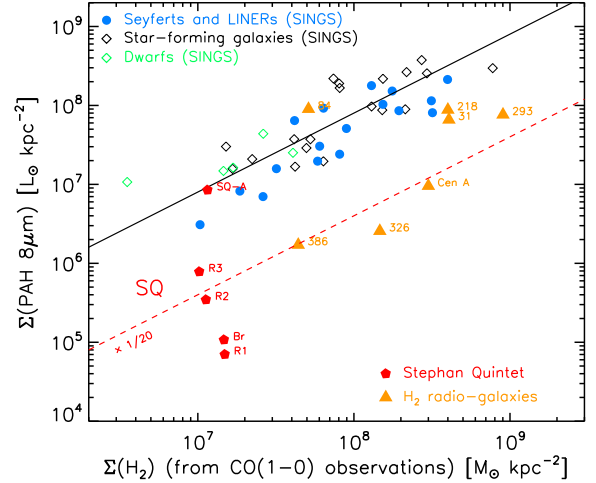}
      \caption{Star-formation rate (assumed to be traced by the surface luminosity of the PAH emission) as a function of the surface mass of  $\rm H_2$, derived from CO(1-0) line measurements, integrated over the full velocity range. The solid line shows the relationship obtained for the SINGS sample. Empty diamonds and filled dots mark dwarf galaxies, star-forming galaxies, and AGN. The SINGS data is from \citet{Roussel2007}. The filled red pentagons show the positions within the Stephan's Quintet group where we performed CO observations and extracted IRS spectra. R1, R2 and R3 are the three positions in the ridge. SQ-A denotes the northern star-forming region and Br the bridge (see Figure~\ref{fig:CO_beams_on_HST}). For comparison, the orange triangles mark the $\rm H_2$-luminous radio galaxies of the \citet{Ogle2010} sample for which we have both clear PAH and CO detections.}
       \label{fig:KSplot_PAH}
\end{figure}

\begin{deluxetable}{l c c c c}
 \tablecaption{Intensities of the PAHs complexes}
 \tablecolumns{5}
 \tablewidth{\columnwidth}
 \tablehead{
 \colhead{Target region} &  \colhead{$7.7\,\mu$m} &  \colhead{$11.3\,\mu$m} &  \colhead{$12.6\,\mu$m} &  \colhead{$17\,\mu$m}  }
\startdata
SQ-A 	  	& 							 & 							  & 							   & $ 4.67 \pm 0.21 $	 \\		
Ridge~1   	& $ 1.54 \pm 1.20 $ & $ 8.53 \pm 0.48 $ & $ 2.23 \pm 0.22 $ & $ 1.84 \pm 0.36 $  \\
Ridge~2  	& $ 7.60 \pm 1.09 $ & $ 7.59 \pm 0.26 $ & $ 3.01 \pm 0.26 $ & $ 2.79 \pm 0.32 $  \\
Ridge~3   	& $ 17.2 \pm 1.44 $ & $ 9.31 \pm 0.27 $ & $ 1.80 \pm 0.37 $ & $ 1.50 \pm 0.27 $  \\
Bridge	  	& 							 & 							  & 							   & $ 0.96 \pm 0.26 $	 
\enddata
\tablecomments{PAH intensities derived from \textit{Spitzer IRS} spectra at the positions of our CO observations, and integrated over the CO(1-0) beam. Fluxes are in units of $10^{-17}$ W~m$^{-2}$.  }
\label{table_PAH_intensities}
\end{deluxetable}

\begin{deluxetable*}{c|ccc|cccc} 
\tablecolumns{8}
\tablewidth{\textwidth} 
\tablecaption{Mass, energy and luminosity budgets of the pre-shock and post-shock gas in the SQ ridge. }
\tablehead{& \multicolumn{3}{c|}{{\sc pre-shock gas}} &  \multicolumn{4}{c}{{\sc post-shock gas}}}
\startdata
gas phases   & Hot Plasma\tablenotemark{a} & H{\sc i}\tablenotemark{b} & $\rm H_2$ \tablenotemark{e} &  Hot Plasma\tablenotemark{a} & H{\sc ii}\tablenotemark{c}  & H{\sc i}\tablenotemark{b} & $\rm H_2$ \tablenotemark{e} \\
\tableline
   $n_{\rm H}$ [cm$^{-3}$] & $3.2 \times 10^{-3}$ & \nodata &\nodata & $1.17 \times 10^{-2}$ & \nodata &\nodata & \nodata \\
  $T$ [K] & $5.7 \times 10^{6}$ &\nodata  & $10-10^{3}$ & $6.9 \times 10^{6}$ &  \nodata &  \nodata  & $10-10^{3}$ \\
  $P / k_{\rm B}$ [ K cm$^{-3}$] & \nodata &\nodata  &\nodata & $1.9 \times 10^{5}$ & \nodata & \nodata & \nodata  \\
   $N_{\rm H}$ [cm$^{-2}$] & \nodata & $3 \times 10^{20}$ & \nodata & \nodata  & $1.4 \times 10^{19}$  & $< 5.8 \times 10^{19}$ & $5 \times 10^{20}$  \\
\tableline
Masses [$\rm M_{\odot}$] & $ 2.6 \times 10^8 $ &  $1.0 - 2.8 \times 10^{9}$ &  $<2.5 \times 10^9$ & $ 5.3 \times 10^{8}$ & $4.9 \times 10^7$ & $< 2 \times 10^8$  & $1.6 - 4.1 \times 10^9$ \\
\tableline
	\multirow{2}*{Energy [erg]} & \multicolumn{1}{c}{Thermal} & \multicolumn{2}{c |}{Kinetic} & Thermal &  & & Kinetic \\
& $3.4  \times 10^{56}$ & $0.5  - 1.3 \times 10^{58}$ &  $<1.9\times 10^{57}$ & $ 9 \times 10^{56}$ & $6 \times 10^{55}$  & \nodata  & $4 - 13 \times 10^{57}$ \\
\tableline
  \multirow{2}*{Flux [W m$^{-2}$]} & \multicolumn{1}{c}{X-rays} & & CO(1-0) & \multicolumn{1}{c}{X-rays} & \multicolumn{1}{c}{H$\, \alpha$} &   \multicolumn{1}{c}{O{\sc i}} & mid-IR H$_{2}$ lines \tablenotemark{d} \\
& $6.6 \times 10^{-17}$ &  \nodata   & $<1.9 \times 10^{-17}$ & $31.4 \times 10^{-17}$ & $7.8 \times 10^{-17}$ & $5.5 \times 10^{-17}$ & $7.5 \times 10^{-16}$  \\
\hline
$\mathcal{L}$\tablenotemark{f}  [erg s$^{-1}$] & $6.2 \times 10^{40}$ &   \nodata  &   $< 2 \times 10^{40}$ & $2.95 \times 10^{41}$ &  $6.5 \times 10^{40}$ &  $4.6 \times 10^{40}$ &  $8 \times 10^{41}$ 
\enddata
    \label{table_mass_NRJ_budgets_SQ}
\tablecomments{All the numbers are scaled to the main shock aperture defined in $\ $ \citet{Cluver2010}: $A_{\rm MS} = 77 \times 30 \ \rm arcsec ^2 = 37.2 \times 13.2$~kpc$^{2}$. This aperture does not include the bridge region. The pre-shock gas is mainly in the form of  H{\sc i} gas, contained in NGC~7319's tidal filament, and molecular gas. After the shock, the mass is distributed between the hot X-ray emitting plasma and the  $\rm H_2$ gas \citep[see discussion in][]{Guillard2009}. }
\tablenotetext{a}{\textit{Chandra} observations of the extended X-ray emission in the shock and the tidal tail \citep{O'Sullivan2009}.}
\tablenotetext{b}{Based on extrapolation of H{\sc i} observations in the tidal tail by  \citet{Sulentic2001, Williams2002}.}
\tablenotetext{c}{H$\, \alpha$ and O{\sc i} optical line observations by  \citet{Xu2003}.}
\tablenotetext{d}{From \textit{Spitzer IRS} observations. The $\rm H_2$ line flux is summed over the S(0) to S(5) lines \citep{Cluver2010}.}
\tablenotetext{e}{Derived from the CO observations presented in this paper, assuming a Galactic conversion factor.}
\tablenotetext{f}{Luminosities are indicated assuming a distance to SQ of 94~Mpc.}
\end{deluxetable*}

Following \citet{Nesvadba2010}, Figure~\ref{fig:KSplot_PAH} presents a diagram similar to the classical Schmidt-Kennicutt relationship. Based on the empirical correlation between PAH emission and star formation in starbursts \citep{Calzetti2007, Pope2008}, the $7.7\,\mu$m PAH emission is used as a tracer of star formation.  The intensities of the PAH complexes measured on the IRS spectra are listed in Table~\ref{table_PAH_intensities}. 

We find that the SQ shock and bridge regions are significantly offset (up to $-1.9$~dex) from the correlation obtained with the SINGs sources \citep[data taken from][]{Roussel2007}. Similarly low PAH-to-CO surface luminosity ratios are also found in some H$_2$-luminous radio galaxies \citep{Nesvadba2010}. 
We interpret this relative suppression of PAH emission as evidence for a lower star formation efficiency.

Part of this offset could come from an under-abundance of PAHs with respect to the Galactic value (possibly caused by their destruction in shocks).  Our modelling of the dust emission showed that the observed PAH and mid-IR continuum emission are consistent with the expected emission from the warm H$_2$ gas in the shock for a Galactic dust-to-gas mass ratio and Galactic PAHs and very small grains (VSGs) abundances \citep{Guillard2010}. Since the H$_2$ mass derived from the CO observations is only a factor of 5 (or less, given the uncertainties on the X-factor) higher than the H$_2$ mass derived from IRS spectroscopy, it is unlikely that the PAH abundance would be depleted by more than an order of magnitude. Therefore, we favour the possibility that the efficiency of conversion from molecular gas to stars in the shock is much lower than in ``normal'' star forming galaxies. On the contrary, the SQ-A region is consistent with these galaxies and with star-forming collision debris in colliding galaxies \citep{Boquien2010}. In this region, the star formation rate is more than one order of magnitude higher than in the shock \citep{Cluver2010}.

Interestingly, we find that the regions where the star formation is the most suppressed are those where the CO velocity dispersion is the highest. 
The very high velocity dispersion observed in the CO gas in the SQ shock and bridge could maintain most of this gas in a diffuse and non-gravitationally bound state, which would explain why this molecular gas is inefficient at forming stars. This trend is also observed in other interacting systems like the tidal dwarf galaxy VCC~2062 \citep{Duc2007} or Arp~94 \citep{Lisenfeld2008}, where the star formation is only observed in regions where the CO lines have a small velocity dispersion. Suppression of star formation in highly turbulent molecular disks 
has also been observed in radio-galaxies \citep{Nesvadba2011a}.

\subsection{Origin of the CO gas in the IGM of SQ: pre- or post-shock gas?}
\label{subsec:origin_CO}

The CO gas in the IGM of SQ has two possible origins: it is either associated with the pre-shock or the post-shock gas \citep[see discussion in][]{Guillard2009}. 

We call \textit{pre-shock} molecular gas the H$_2$ clouds that have not seen the fast ($\approx 600$~km~s$^{-1}$) intercloud shock produced by the galaxy collision (materialized by the X-ray emitting ridge), or the clouds for which the fast intercloud shock have swept across them, but these clouds have survived to the shock. In the latter case, these pre-shock molecular clouds must be dense enough not to be ionized by the radiation field of the fast intercloud shock (propagating in the tenuous, $n_{\rm H} \approx 10^{-2}$~cm$^{-3}$, and X-ray emitting gas). Such clouds have too much inertia to be decelerated by the tenuous gas, and  retain their pre-shock velocity (in Appendix we show that the deceleration timescale is longer than the age of the shock). Therefore, we expect to detect this CO gas at velocities close to the two pre-shock velocities ($5700-6000$ and $6700$~km~s$^{-1}$). 

We call \textit{post-shock} gas the molecular gas that is formed from diffuse gas (e.g. H{\sc i} gas) that has been heated by the transmitted shocks into the clouds, decelerated in the frame of reference of the intercloud shock front, and that had time to cool and become molecular within this deceleration timescale. This is the scenario that \citet{Guillard2009} proposed to explain the formation of H$_2$ in the SQ ridge. 
This post-shock H$_2$ gas is dynamically coupled to the lower density phases. In the frame of reference of the observer, the post-shock CO gas, formed from shocked intruder gas, is decelerated to velocities redder than the pre-shock intruder velocity (at $V \gtrsim 5700-6000$~km~s$^{-1}$), while the post-shock CO gas, formed from shocked IGM gas\footnote{CO should form on a similar timescale as $\rm H_2$ in the post-shock gas \citep[see Figure~C.2 of][]{Guillard2009}.}, is decelerated to velocities bluer than the pre-shock IGM velocity ($V \lesssim 6700$~km~s$^{-1}$). Therefore, some post-shock CO gas should appear at a velocity intermediate to that of the two colliding gas flows. 
However, depending on the coupling efficiency and the dynamics of the galaxies in the group, some post-shock gas could also be found close to the IGM (NGC~7319's tidal debris) or the intruder (NGC~7318b) velocities.

The detection of CO emitting gas at both pre-shock and intermediate velocities between that of the intruder and that of the IGM   
shows that the molecular gas in the IGM of SQ originates from both pre-shock clouds, and from the formation of molecular gas in the cooling post-shock gas that has been accelerated in the frame of the shock.
The observed CO kinematics show that the pre-shock components have lower velocity dispersions than the post-shock components (sect.~\ref{subsec:high_CO_vel_disp}), suggesting that the shock is driving turbulence into the post-shock gas.

\subsection{Pre- and post-shock gas masses and energy budgets}
\label{subsec:mass_nrj_budgets}

The mass and energy budget of the pre-shock and post-shock gas in the ridge is summarized in Table~\ref{table_mass_NRJ_budgets_SQ}, which is an updated version of  Table~1 in \citet{Guillard2009}, now taking into account the CO data, and the H$_2$ spectral mapping of the shock (ridge) region by \citet{Cluver2010}. 
Distinguishing spatially between the pre- and post-shock gas is difficult with the present CO observations. We rather rely on the kinematics to separate and quantify the masses and kinetic energies carried by each component.  The uncertainties in the relative masses associated with both components are high, essentially because the dynamical coupling of the H$_2$ gas with lower density phases and the global dynamics of the system are poorly constrained. In Table~\ref{table_mass_NRJ_budgets_SQ},  the CO components with central velocities in the ranges $5700-6100$ and $6600-6800$~km~s$ ^{-1} $ are associated with the pre-shock gas of the intruder and the intra-group medium, according to H{\sc i} \citep{Sulentic2001} and optical data \citep{Xu2003}. Outside these ranges, the CO emission is counted as post-shock gas. These velocity boundaries are highly uncertain and are likely to underestimate the proportion of post-shock gas, and over-estimate the pre-shock gas masses. We thus consider the post-shock gas masses as lower limits, and pre-shock gas masses as upper limits. In particular we choose to include the 6000~km~s$^{-1}$ component in the pre-shock budget as suggested in \citet{Moles1997, Sulentic2001, Xu2003}, but this is questioned by other authors \citep{Williams2002, Lisenfeld2002} who argue this is intruder gas that has been accelerated by the shock from 5700 to 6000~km~s$^{-1}$. 

Under this assumption, the post-shock cold H$_2$ mass is $\approx 1/3$ of the mass detected at the pre-shock velocities. Alternatively, if we consider that the 6000~km~s$^{-1}$ component belongs to the post-shock gas,  this proportion is $\approx 3/2$. In either case, both pre-shock origin and post-shock CO formation are important. 

The kinetic energy associated with pre- or post-shock gas has been calculated from Eq.~\ref{eq:Ekin_int}, based on the same velocity separation as for the masses (Sect.~\ref{subsec:mass_nrj_budgets}). For the post-shock gas, we quote the range of kinetic energy, being computed velocity component by component, or integrated over the full velocity range (after subtraction of the CO signal at the pre-shock velocities).

\section{Conclusions and final remarks}
\label{sec:conclusion}

We report on CO(1-0), (2-1) and (3-2) line observations with the IRAM 30m and APEX telescopes in the galaxy-wide shock of Stephan's Quintet, aimed at studying the CO gas counterpart to the powerful warm H$_2$ emission detected by \textit{Spitzer}/IRS. The main results are the following:
\begin{itemize}

\item
Multiple CO velocity components are detected, at the pre-shock velocities (namely that of the intruder galaxy at 5700~km~s$^{-1}$ and that of the intra-group medium at 6700~km~s$^{-1}$), but also at intermediate ($6000-6400$~km~s$^{-1}$), and higher (6900~km~s$^{-1}$ in SQ-A) velocities. 
The CO gas originates from both pre-shock molecular gas, and post-shock gas, i.e. diffuse gas heated by the shock that had time to cool and become molecular within the collision age. CO lines are intrinsically broader in the ridge and in the bridge than in the SQ-A and southern star-forming regions. A large fraction of the gas kinetic energy of the collision is carried by the turbulent velocity dispersion of the CO emitting gas, which means that most of this energy has not been thermalized yet.

\item 
The CO(1-0) and (2-1) emission is distributed over the intra-group shock (the ridge) and in an eastern bridge feature that connects the shock to NGC~7319. We also detected the CO(3-2) line in the ridge and tentatively in SQ-A. Large amounts of CO emitting gas are detected, $\approx 5.4\times 10^9\,$M$_{\odot}$ (after conversion from CO(1-0) emission to H$_2$ using the Galactic value) in the ridge and in the bridge. Note that this mass could be lower by a factor of a few because of the large uncertainties on the CO to H$_2$ conversion factor. This gas is co-spatial  with warm $\rm H_2$ seen in the mid-IR domain, H$\,${\sc ii} gas, and X-ray emitting hot plasma. The ratio  between the H$_2$ mass derived from the \textit{Spitzer} spectroscopy to the H$_2$ mass derived from the CO observations is $\approx 0.2$, which is $1-2$ orders of magnitude higher than in star-forming galaxies, showing that the heating rate of the molecular gas in SQ is particularly high. This high heating rate is likely to be powered by the turbulent dissipation of the kinetic energy involved in the galaxy collision. This interpretation implies that the velocity dispersion on the scale of giant molecular clouds (GMCs)  in SQ is an order  of magnitude larger than the Galactic value.

\item
We find that the ratio of the PAH-to-CO surface luminosities in the shock and in the bridge regions are much lower (up to a factor 75) than the observed values in star-forming galaxies. In the SQ-A star-forming region, the PAH-to-CO ratio is compatible with the classical Schmidt-Kennicutt law.
We suggest that the star formation efficiency  in the shock of SQ is lower than in ``normal'' galaxies. This could be a consequence of the high turbulence within the H$_2$ gas. 

\end{itemize}

Our census of the molecular gas in the IGM of SQ is still incomplete. High angular resolution interferometric radio observations of SQ will help to determine the clumpiness of the CO emitting gas and perhaps spatially separate the different velocity components. Far-infrared photometry is also needed to estimate the total dust mass, and thus the cold gas content independently of the CO to H$_2$ conversion factor.

A global model for the formation and kinematics of the molecular gas present in the IGM of SQ can only be addressed by means of complex numerical simulations. A first step has been achieved by \citet{Renaud2010}, who presented a pilot $N$-body study with a solution that matches the overall geometry of the group. 
Hydrodynamical simulations by \citet{Hwang2011} took a step further including the gas component and made first  headways in our understanding of the gas kinematics in the system. However, presently, the existing numerical simulations do not include the molecular gas and cannot reproduce the CO and H$_2$ observations of SQ presented here and in \citet{Appleton2006, Cluver2010},  where most of the molecular gas is not associated with star formation. The prescriptions used in the numerical models to describe the conversion from interstellar gas to stars will have to be refined, especially in the treatment of the gas cooling, energy exchanges between the phases of the interstellar medium, and turbulent cascade of the kinetic energy of the collision.

\acknowledgements

Based on observations carried out with the IRAM 30-meter telescope. IRAM is supported by INSU/CNRS (France), MPG (Germany) and 
IGN (Spain). PG also would like to acknowledge in particular the IRAM 
staff for help provided during the observations. 
UL acknowledges support  by the research project 
AYA2007-67625-C02-02 from the Spanish Ministerio de Ciencia y
Educaci\'on and the Junta de Andaluc\'\i a (Spain) grant FQM-0108.

Part of this publication is based on data acquired with the Atacama Pathfinder EXperiment (APEX). APEX is a collaboration between the Max-Planck-Institut f\"ur Radioastronomie, the European Southern Observatory, and the Onsala Space Observatory.

This work is partly based on observations made with the \textit{Spitzer Space Telescope}, which is operated
by the Jet Propulsion Laboratory, California Institute of Technology, under a contract with NASA. 

This research has made use of the NASA/IPAC Extragalactic Database (NED)
which is operated by the Jet Propulsion Laboratory, California Institute of Technology, under
contract with the National Aeronautics and Space Administration.

\bibliographystyle{apj}
\bibliography{Guillard_CO_SQ.bbl}

\appendix

\subsection{Dynamical coupling of the CO gas to other gas phases}
\label{subsec:dyn_coup_CO}

In the following we show that it is unlikely that the CO gas observed at intermediate velocities originate from pre-shock molecular clouds, essentially because the timescale of dynamical coupling, i.e. the time it takes for the pre-shock cloud to decelerate and become comoving with the shocked intercloud medium is much longer than the age of the collision. 

Let us consider the case of an H$_2$ cloud of density $\rho _c$ and radius $R_c$, initially in pressure equilibrium with the inter-cloud medium (of density $\rho _i$), that is run over by a fast shock wave of velocity $V_i$ in the inter-cloud medium. The timescale for the cloud compression due to the rise of the external pressure, so-called the \textit{crushing time}, is given by $t_{\rm crush} = R_c/V_c$, where $V_c \simeq \sqrt{\frac{\rho _i}{\rho _c}} V_i = V_i / \sqrt{\chi}$ is the velocity of the transmitted shock into the cloud ($\chi = \rho _c / \rho _i $ is the density contrast between the two media).
The cloud acceleration timescale, $t_{\rm acc}$, i.e. the amount of time it takes to accelerate the cloud  to the velocity of the post-shock background $V_{\rm ps, i} = \frac{3}{4} V_i$ is \citep[e.g.][]{Klein1994}:
\begin{eqnarray}
\label{eq:cloud-acceleration-timescale}
t_{\rm acc} & =& \frac{16}{9} \,  \frac{\rho _{\rm c}}{\rho _i} \,   \frac{R_{\rm c}}{ V_{\rm i}} =  \frac{16}{9}  \,   \sqrt{\chi} \,   t_{\rm crush} \\
& \simeq & 1.7 \times 10^8 \left( \frac{\chi}{10^3} \right) \left( \frac{R_{\rm c}}{10\,\rm pc}\right) \left( \frac{V_{\rm i}}{100\,\rm km\,s^{-1}}\right) ^{-1} \, \rm [yr].
\end{eqnarray}
The above equation show that it takes a long time to couple dynamically pre-shock dense H$_2$ gas with the background flow of tenuous gas. The cloud crushing time (time for the inter-cloud shock to go through the cloud) to acceleration time ratio is $t_{\rm crush} / t_{\rm acc}  = 9 / 16 \sqrt{\chi}$, which shows that as soon as the density contrast is higher than $\approx 10^2$, the cloud will be compressed on a timescale much shorter than the acceleration time. 
In the SQ shock, the high density contrast between cold clouds and the hot tenuous flow \citep[$\chi > 10^5$, see][]{Guillard2009} would lead to deceleration timescales longer than the shock age.

On the other hand, in the model of $\rm H_2$ formation in the shock proposed by \citet{Guillard2009}, the molecular gas is formed out of moderate density medium ($n_{\rm H} \approx 1$~cm$^{-3}$) that is compressed and decelerated while becoming molecular. 
Therefore, in this context, the coupling is much more efficient because the density contrast is much lower ($\chi < 10^2$).
This \textsl{in-situ} formation is  a solution to explain the high velocity dispersion seen in the spectra, although we do not exclude the presence of pre-shock dense clouds.

\end{document}